\documentclass[a4paper, twocolumn, accepted=2024-04-24]{quantumarticle}
\pdfoutput=1
\usepackage[utf8]{inputenc}
\usepackage[english]{babel}
\usepackage[T1]{fontenc}
\usepackage{amsmath}
\usepackage{tikz}
\usepackage{lipsum}
\usepackage[colorlinks=true,linkcolor=blue,citecolor=blue]{hyperref}
\usepackage{adjustbox}
\usepackage{ifthen}
\usepackage{forest}
\usepackage{nicefrac}
\usepackage{braket}
\usepackage{tikz}
\usepackage{dsfont}
\usepackage{amsthm}

\usepackage{bm}
\usepackage[mode=buildnew]{standalone}% requires -shell-escape
\usepackage{amsmath}
\usepackage{amssymb}
\usepackage{physics}
\usepackage[english]{babel}
\usepackage{graphicx}
\usepackage{colortbl}
\usepackage{xcolor}
\usepackage{soul}
\usepackage{algpseudocode}
\usepackage{algorithm}
\usepackage[version=3]{mhchem}
\usepackage{physics}
\usepackage{siunitx}
\usepackage{algorithm}
\usepackage{algpseudocode}
\usepackage{tikz}
\usepackage{smartdiagram}
\usepackage{setspace}
\usepackage{float}
\usepackage[numbers]{natbib}

\usetikzlibrary{positioning}
\tikzset{main node/.style={circle,fill=blue!20,draw,minimum size=1cm,inner sep=0pt},
}
\usetikzlibrary{overlay-beamer-styles}
\usetikzlibrary{shapes.geometric}
\usetikzlibrary{positioning}
\usetikzlibrary{overlay-beamer-styles}
\usetikzlibrary{shapes,arrows,shadows,positioning}
\usetikzlibrary{arrows, decorations.markings}
\usetikzlibrary{backgrounds}
\usetikzlibrary{decorations.pathreplacing}
\usetikzlibrary{3d,decorations.text,shapes.arrows,positioning,fit,backgrounds}
\usepackage{tikz-network}
\usetikzlibrary{overlay-beamer-styles}
\usetikzlibrary{shapes.geometric}

\algnewcommand{\LineComment}[1]{\State \(\triangleright\) #1}

\begin{document}
\title{Hybrid discrete-continuous compilation of trapped-ion quantum circuits with deep reinforcement learning}
\author{Francesco Preti}
\affiliation{Forschungszentrum J\"ulich, Institute of Quantum Control (PGI-8), D-52425 J\"ulich, Germany}
\affiliation{University of Cologne, Institute of Theoretical Physics, , D-50937 K\"oln, Germany}
\author{Michael Schilling}
\affiliation{Forschungszentrum J\"ulich, Institute of Quantum Control (PGI-8), D-52425 J\"ulich, Germany}
\affiliation{University of Cologne, Institute of Theoretical Physics, , D-50937 K\"oln, Germany}
\author{Sofiene Jerbi}
\affiliation{University of Innsbruck, Institute for Theoretical Physics, A-6020 Innsbruck, Austria}
\affiliation{Freie Universit\"{a}t Berlin, Dahlem Center for Complex Quantum Systems, D-14195 Berlin, Germany.}
\author{Lea M. Trenkwalder}
\affiliation{University of Innsbruck, Institute for Theoretical Physics, A-6020 Innsbruck, Austria}
\author{Hendrik Poulsen Nautrup}
\affiliation{University of Innsbruck, Institute for Theoretical Physics, A-6020 Innsbruck, Austria}
\author{Felix Motzoi}
\affiliation{Forschungszentrum J\"ulich, Institute of Quantum Control (PGI-8), D-52425 J\"ulich, Germany}
\affiliation{University of Cologne, Institute of Theoretical Physics, , D-50937 K\"oln, Germany}
\author{Hans J. Briegel}
\affiliation{University of Innsbruck, Institute for Theoretical Physics, A-6020 Innsbruck, Austria}
\date{03-05-2024}

\begin{abstract}
Shortening quantum circuits is crucial to reducing the destructive effect of environmental decoherence and enabling useful algorithms. Here, we demonstrate an improvement in such compilation tasks via a combination of using hybrid discrete-continuous optimization across a continuous gate set, and architecture-tailored implementation. The continuous parameters are discovered with a gradient-based optimization algorithm, while in tandem the optimal gate orderings are learned via a deep reinforcement learning algorithm, based on projective simulation. To test this approach, we introduce a framework to simulate collective gates in trapped-ion systems efficiently on a classical device. The algorithm proves able to significantly reduce the size of relevant quantum circuits for trapped-ion computing. Furthermore, we show that our framework can also be applied to an experimental setup whose goal is to reproduce an unknown unitary process.  
\end{abstract}

\maketitle

\section{Introduction}
The last decade has seen significant progress in the development of quantum computing architectures \cite{Arute2019}. While scalable fault-tolerant quantum computers are still out of reach in the near future, noisy, intermediate-scale quantum (NISQ) computers may already offer some benefits over classical ones for specific computational tasks \cite{Preskill2018quantumcomputingin, Bravo-Prieto2019}. In particular, variational algorithms \cite{Peruzzo2014, Cerezo2021review}, where most of the operations depend on several continuous parameters, have emerged as a suitable class of methods that could potentially achieve quantum speed-up on NISQ devices. 

Implementing a high-level quantum algorithm both on fault-tolerant and NISQ devices requires adequate methods to compile it in the set of universal quantum gates available to the hardware. While several frameworks for compilation of quantum circuit-based algorithms on physical platforms are being developed \cite{Kreppel2022, BQSKit, Qiskit}, common available approaches, such as heuristic and automated search \cite{venturelli2019quantum, Wille2020}, in many case cannot output an optimal circuit for a specific target operation \cite{Maronese2021}.

In digital quantum computers, compilers implement a general quantum circuit through a discrete set of universal quantum gates. In real physical quantum devices, however, and more generally in analog computation, an additional layer of complexity is present, due to the necessity of optimizing continuous gate parameters to reproduce target unitaries. These parameters may depend, e.g., on the specific Hamiltonian employed in the quantum computing platform, such as the XY-Hamiltonian or the Mølmer–Sørensen interaction for trapped ions \cite{Soerensen1999_1, Soerensen1999_2}, the cross resonance interaction for IBM quantum computers \cite{blogCRexplained} or the Fermi-Hubbard model for neutral atoms \cite{DallaireDemers2016}.
As a result, when taking into account physical parameters, variational algorithms, and generally continuous gate sets \cite{Preti2022}, one must supplement the circuit compilation task with a subsequent optimization of the parameters defining the individual constituent gates. For the optimization of continuous parameters, we have several options available, e.g., gradient-based algorithms
\cite{Daoud2022}, evolutionary algorithms \cite{Simon2013} and direct search \cite{NelderMead1965}. For the compilation of the circuit gate structure, a standard approach is given by methods based on the Solovay-Kitaev algorithm \cite{dawson2005solovay}, different circuit factorization strategies \cite{Farrokh2004, Wille2016}, graph path traversal algorithms, such as the $A^*$ algorithm \cite{Davis2020}, semi-definite programming and and various machine learning methods, including reinforcement learning  \cite{Maronese2021}. More specifically, deep reinforcement learning has been recently successfully implemented for the optimization of discrete quantum circuits \cite{Moro2021, Foesel2021, Ostaszewski2021, Sivak2022, Borah2021, Zhang2020, yao2021reinforcement}.

In reinforcement learning (RL) \cite{Sutton1998}, an agent learns to maximize a properly engineered reward signal by interacting with an environment, which encodes the optimization task to solve. RL has already been applied to solve various challenging tasks, including, e.g., surpassing human performance in certain classes of games \cite{Silver2017} or in complex, computationally expensive problems such as protein folding \cite{Jumper2021}. Projective Simulation (PS) is a physics-inspired framework for intelligent agents which has also been applied to solve RL tasks in quantum physics  \cite{Briegel2012, Melnikov2017, Melnikov1221, Wallnfer2019MachineLF,dalgaard2020global, Nautrup2019optimizingquantum} and biology \cite{ried2017modelling}. This model can naturally be extended to a deep RL model \cite{Jerbi2021, Preti2020} and has found applications in representation learning \cite{Nautrup2020, Eva2022}.

In this work, we propose a unified approach to optimizing the placement and parameter optimization of the gates in the circuit. We argue that such an approach is both relevant to traditional compilation of a more expressive, continuous gateset \cite{Preti2022}, as well as to variational circuits where the experimental cost-function optimization may be simultaneously performed over discrete and continuous degrees of freedom. We use a RL agent, based on the PS framework, for the combinatorial optimization and a gradient-based optimizer for the continuous optimization of the gate angles. In particular, we extend the method proposed in \cite{Ostaszewski2021} for the optimization of variational circuits in quantum chemistry and \cite{Sivak2022} for state preparation to the case of unitary compilation. We consider the framework where an agent, that has control over an experimental platform, interacts with a black-box unitary process and attempts to simulate it. The task of the RL agent is to optimize the position of the gates on the circuit, whereas the gradient-based optimizer finds the optimal set of continuous parameters that minimize a given cost function. The reward function for the RL agent is constructed based on the results of the continuous optimization. 

We test the learning algorithm first on standard unitaries, such as Toffoli gates, and then consider the task of quantum process simulation. The latter can be conceived as an experimental black-box unitary approximation strategy that allows the agent to reconstruct an unknown unitary process by compiling a proper quantum circuit. 

Independently of our circuit compilation results, we propose a method to speed-up the simulation of quantum circuits based on an efficient representation of trapped-ion gates replacing standard matrix exponentiation. This method allows us to obtain analytic expressions for the ion-trap gates and their gradients and also to simulate the given gate set on other quantum computing platforms, such as superconducting quantum circuits \cite{Kwon2021} or neutral atoms \cite{Saffman2016}.

The paper is organized as follows: In Section~\ref{sec:gateset} we introduce the quantum circuit framework for trapped-ions. In Section~\ref{sec:dynamics} we present our method to compute fast analytic ion gates in simulation. In Section~\ref{sec:optim} we discuss continuous optimization methods and strategies to compute the gradients of the cost function both in simulation and on a real quantum device. In Section~\ref{sec:rl} we introduce PS and its extensions in the context of RL methods. In Section~\ref{sec:circ_comp} we introduce the problem of circuit synthesis and our hybrid RL-continuous optimization method. In Section~\ref{sec:results} we discuss the results of applying the proposed method to the compilation of (black-box) unitaries.

\section{Problem setting}\label{sec:gateset}
In this work, we consider a specific set of gates, normally implemented in trapped-ion quantum circuits, which is based on global Mølmer–Sørensen (MS) gates, equatorial  rotations acting on the entire register of qubits, as well as local polar rotations. Trapped ions are among the most promising platforms for quantum computing hardware \cite{Bruzewicz2019}. They exhibit impressive coherence times even in absence of dynamical decoupling and spin echo techniques \cite{Harty2014} and have been shown to allow for high-fidelity quantum gates \cite{Ballance2016, Erhard2019}. In a trapped-ion quantum computer, ions -- usually \ce{^{43}Ca^+} -- are confined in a Paul trap \cite{Paul1990} using a varying electromagnetic field. The ions are addressed individually by a system of lasers aligned externally to the trap. The interaction of the laser field with the ion motional and electronic degrees of freedom allows for entangling operations. 
The laser pulses can be engineered to define the following universal set of quantum gates \cite{Martinez2016, Gaebler2012, Monroe2014}:
\begin{align}
	&\text{MS}(\theta, \phi) = \exp{-i  \frac{\theta}{4}( S_x  \cos \phi+  S_y \sin \phi)^2} \label{eq: ms gate set} \\
	&C_{xy}(\theta, \phi)= \exp{-i \frac{\theta}{2}( S_x  \cos \phi+  S_y \sin \phi)} \label{eq: ms gate set2} \\
	&Z_j(\theta) = \exp{-i\frac{\theta}{2} \sigma_z^{(j)}} \label{eq: ms gate set3},
\end{align}
where the first gate is a global MS gate \cite{Soerensen1999_1, Soerensen1999_2}, the second gate is a rotation of the entire register of qubits in the equatorial plane of the Bloch sphere, and the third gate represents a single-qubit, and therefore local, $\sigma_z$ rotation acting on qubit $j$. 
The operators $S_x = \sum_{i=1}^{n} \sigma_x^{(i)}$, $S_y = \sum_{i=1}^{n} \sigma_y^{(i)}$,  $S_z = \sum_{i=1}^{n} \sigma_z^{(i)}$ are given by the Pauli operators acting on qubit $i$. 
These gates can be easily generalized to the qudit case \cite{Ringbauer2022}. For the optimization of unitaries, the gate overlap fidelity is a standard figure of merit \cite{Nielsen2002}:
\begin{align}\label{eq:gatefid}
    F(U,V) = \frac{1}{d^2} \vert \tr{VU^{\dagger}} \vert ^2,
\end{align}
where $U$ and $V$ are unitaries (one of them is the goal of the optimization) and $d$ is the dimension of the Hilbert space -- for $n$-qubit systems $d=2^n$. Commonly, synthesising quantum circuits to reproduce an arbitrary unitary $U$ requires having access to quantum computing hardware that implements a universal gate set and running the optimization of the continuous parameters. 

\begin{figure*}[ht!]
	\hspace*{-1cm}
	\centering
	\includegraphics[width=16cm]{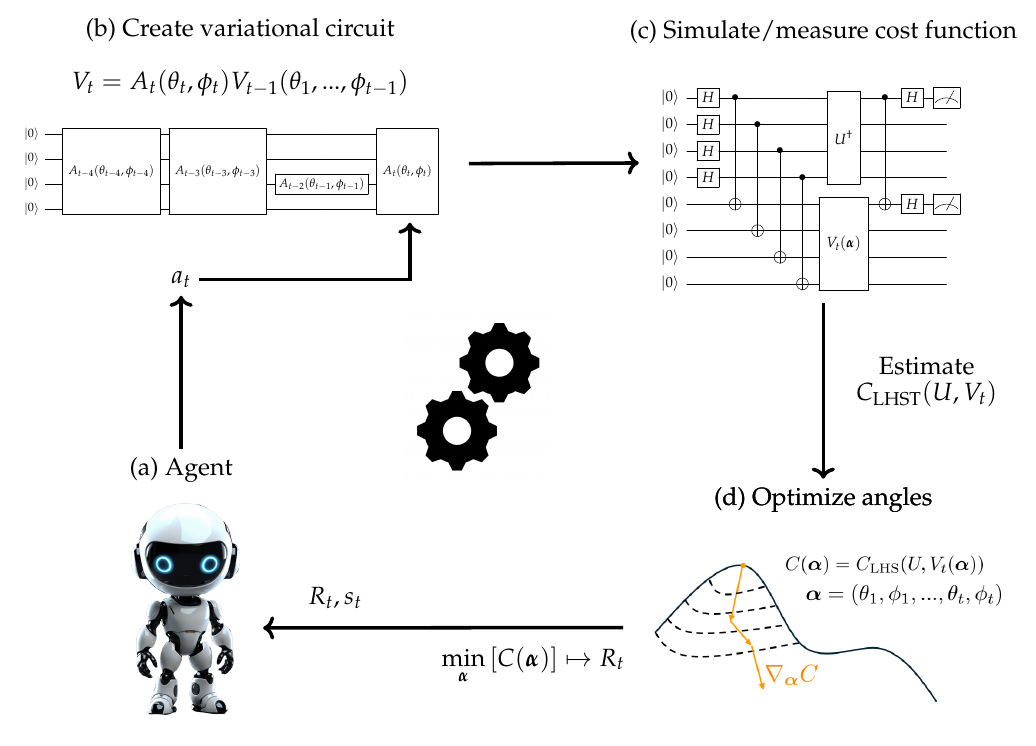}
	\caption[Main RL]{General scheme of the proposed hybrid training loop, where the RL agent (a) learns to optimize a  (variational) quantum circuit. By choosing an action $a_{t} \in \{1, 2, ..., n + 2\}$ the gate $G_{a_{t}}$ is placed on the circuit (b), where $G_1=\text{MS}, G_2=C_{xy}, G_3=Z_1, ..., G_{n+2}=Z_n$ correspond to the gate set introduced in Eqs.~\eqref{eq: ms gate set}-\eqref{eq: ms gate set3}. At each RL time step $t$, the circuit is used to compute a cost function $C(\boldsymbol{\alpha})$ based on the fidelity -- see Eq.~\eqref{eq:gatefid} -- either through classical simulation or the Hilbert-Schmidt test (c) -- see Eq.~\eqref{eq:quantumcost_2} -- experimentally.  The continuous optimizer (d) then outputs a guess for the optimal parameters $\boldsymbol{\alpha}^* = \text{argmin}_{\boldsymbol{\alpha}}\left[ C(\boldsymbol{\alpha})\right]$ that minimize the cost function for a specific circuit $V_{t}$. The minimal value of the cost function is used to assign a reward to the PS agent -- see Eq.\eqref{eq:reward} --, thus closing the RL training loop.}
	\label{fig:rl_ion_diag}
\end{figure*}

If the circuit synthesis is performed on a classical computer, it is also necessary to simulate the gate set efficiently. Simulating and optimizing $n$-qubit collective gates such as those given in Eq.~\eqref{eq: ms gate set} and Eq.~\eqref{eq: ms gate set2} is generally considered more challenging than with just two-qubit entangling gates and single qubit rotations \cite{Martinez2016}. A standard strategy is to progressively increase the number of entangling MS gates on the circuit, accompanied by a suitable number of single and multi-qubit rotations, and to progressively optimize the gate parameters until an acceptable threshold of the figure of merit is reached. This is a viable strategy to obtain one solution for a quantum compilation problem on a trapped-ion device; it is however sub-optimal with respect to the number of gates required. More efficient solutions exist, but the optimization landscape may be difficult to navigate for various algorithms \cite{Dalgaard2022}. In particular, as the optimization landscape with respect to the gate angles is particularly vast and depends on the arrangement of the gates on the circuit, it is often necessary to search through several combinations of gate sequences. Random or automated search can be implemented to reduce the number of gates present on the circuit by arbitrarily trying several configurations \cite{Moro2021, Martinez2016}.

\section{Methods}
In the following section, we discuss our approach to address the three relevant aspects that characterize a circuit synthesis task: the efficient computer-aided simulation of the relevant trapped-ion gates, the optimization of the continuous gate parameters and the optimal arrangement of the gates on the circuit. In addition, we address the possible implementation of our hybrid RL-gradient based optimization method on real quantum devices.

\subsection{Dynamics via fast exponentiation}\label{sec:dynamics}

In simulation, direct computation of the gates in Eqs.~\eqref{eq: ms gate set}-\eqref{eq: ms gate set2} is commonly done via direct matrix exponentiation of a Hamiltonian, which is generally slow for large matrices, since it requires several matrix multiplications, each one with a practical complexity of $O(d^{2.8})$ \cite{Strassen1969}, where $d$ is the matrix dimension. Instead, here we show how to significantly reduce the per-iteration computational cost of the matrix exponentiation by factorizing it with respect to the rotation angle $\theta$ and to the phase angle $\phi$. This is achieved via a spectral decomposition where the phase-independent part can then be cached before the optimization.

Mathematically, the factorization of the matrix exponential for a MS or $C_{xy}$ gate can be written via spectral decomposition as 
\begin{align}\label{eq:xy_exp_main}
    &U_H(\theta, \phi) = \exp{i H(\theta, \phi)} = \\ &=V(\theta, \phi)\left(\sum_{l = 0}^{d-1}  e^{-i\lambda_l(\theta, \phi)} \ketbra{l}{l}\right) V^\dagger (\theta, \phi), \nonumber
\end{align}
where $H(\theta, \phi)$ is a $\theta$- and $\phi$-dependent Hamiltonian -- see exponents in Eqs.~\eqref{eq: ms gate set}-\eqref{eq: ms gate set2} -- and $V$ is the matrix of eigenvectors with respective eigenvalues $\lambda_l$. 

We regroup in terms of degenerate eigenvalues and consider the case where the set of eigenvalues are $\phi$-independent, i.e., $\lambda_l=\lambda_l(\theta)$, while a $\phi$-dependent unitary is applied to the Hamiltonian, i.e., $H(\theta, \phi) = V(\phi)H(\theta)V^{\dagger}(\phi)$. As a consequence, the eigenvectors matrix $V=V(\phi)$ is independent of $\theta$, and we can rewrite
\begin{align}
    U_H(\theta, \phi) =& \sum_{k = 0}^{n_\lambda} V(\phi)\left(  e^{-i\lambda_k(\theta)} \sum_{l_k}\ketbra{l_k}{l_k}\right) V^\dagger (\phi) \\
    =& P(\phi)\odot  \sum_{k = 0}^{n_\lambda} e^{-i\lambda_k(\theta)} D_l,
\end{align}
where $n_\lambda$ is the number of distinct eigenvalues, $\ket{l_k}$ are the respective eigenvectors, $D_k = V (0)\sum_{l_k}\ketbra{l_k}{l_k} V^\dagger(0)$ and $P(\phi)$ is a matrix of phase components. In our case, the columns of $P(\phi)$ are all equal and are given by the vector $\boldsymbol{p}(\phi) = \begin{pmatrix} 1 \\ e^{i \phi} \end{pmatrix}^{\otimes n}$, while $\odot$ represents element-wise (Hadamard) matrix multiplication.
The $C_{xy}$ and MS gates have few unique eigenvalues with $\lambda_k=(2k-n)\theta,\ 0 \leq k \leq n_\lambda=n $ and $\lambda_k=(2k-n)^2\theta,\ 0 \leq k \leq n_\lambda = \lceil n + \frac{1}{2} \rceil $, respectively, making the computation with cached $D_k$ particularly efficient. 
A detailed derivation, together with a discussion of the computational speedup of this representation, is available in Appendix \ref{sec:fast_gates}.

\subsection{Continuous gradient-based optimization}\label{sec:optim}

We consider the optimization of an observable with respect to the continuous parameters. Although single problems can be optimized effectively by implementing an appropriate discretization method, in general this approach can lead to a sub-optimal solution, since it introduces discontinuities in the search space. Instead, we want to consider the dependency of the fidelity from continuous parameters and calculate its gradient.

\subsubsection{Cost function based on known target unitaries.}

We consider a quantum circuit composed of a sequence of continuous gates with angle parameters $\boldsymbol{\alpha} = (\boldsymbol{\alpha}_{1}, ..., \boldsymbol{\alpha}_{L})$ and where each gate is composed of multiple rotation angles $\boldsymbol{\alpha}_{1} \in \mathbb{R}^{M_1}, \ldots, \boldsymbol{\alpha}_{L} \in \mathbb{R}^{M_L}$ and $M_1, ..., M_L \geq 1$ and $M=\sum_{m=1}^L M_m$. The circuit is then given by
\begin{align}\label{eq:general-circuit}
V(\boldsymbol{\alpha}) = \prod_{l=1}^L V_l(\boldsymbol{\alpha}_{l}),
\end{align}
where $V_l(\boldsymbol{\alpha}_{l})$ is an arbitrary parametric unitary with parameters $\boldsymbol{\alpha}_{l}$.

The gradient with respect to a figure of merit, here the average gate fidelity -- see Eq.~\eqref{eq:gatefid} --, can also be directly computed and are given by
\begin{align}\label{eq:grad_v_alpha}
&\nabla_{\boldsymbol{\alpha}_{l}} F = \frac{2}{d^2} \Re{\tr( \nabla_{\boldsymbol{\alpha}_{l}} V(\boldsymbol{\alpha}) U^{\dagger}) \tr(V(\boldsymbol{\alpha})^{\dagger} U)}, 
\end{align}
with
\begin{align}
&\nabla_{\boldsymbol{\alpha}_{l}} V(\boldsymbol{\alpha}) =  V_1(\boldsymbol{\alpha}_{l1}) \cdots \nabla_{\boldsymbol{\alpha}_{l}} V_l(\boldsymbol{\alpha}_{l}) \cdots V_L(\boldsymbol{\alpha}_{L}),
\end{align}
for $1 \leq l \leq L$.
The naive element-wise computation of the gradient components uses $M \cdot L$ matrix multiplications, as the matrix in Eq.~\eqref{eq:grad_v_alpha} needs to be evaluated $M$ times and is given by the product of $L$ unitaries. 
However, the gradient can be computed recursively, a method which is often referred to as GRAPE \cite{Khaneja2005}, by storing the values of the intermediate unitaries $W_l = W_{l-1}V_l^{\dagger}(\boldsymbol{\alpha}_l), W_0 = I_d$ for $l=1,..., L$ in the product
\begin{align}\label{eq:recursive-grad}
\nabla_{\boldsymbol{\alpha}_{l}} F = \frac{2}{d^2}\Re{\tr(W_{l-1} \nabla_{\boldsymbol{\alpha}_{l}} V_l(\boldsymbol{\alpha}_{l})  W_l^{\dagger} VU^{\dagger}) \tr(V^{\dagger} U)}.
\end{align}

The gradient computation method given in Eq.~\eqref{eq:recursive-grad} computes first the intermediate unitaries, for which we need $L$ matrix multiplications and uses them to evaluate the gradient, which compared to Eq.~\eqref{eq:grad_v_alpha} needs only $4M + L + 1$ matrix multiplications and therefore scales linearly with the number of gates and gate parameters in the circuit.

Eq.~\eqref{eq:recursive-grad} allows us to compute the gradient of any cost function that resembles the structure of Eq.~\eqref{eq:gatefid} efficiently in numerical simulations. This can be further sped up by similarly analytically computing the Hessian matrix of the cost function \cite{dalgaard2020hessian} or by using GPU or TPU architectures \cite{Morningstar2022}. Eq.~\eqref{eq:recursive-grad} can also be useful in an experimental setting when the target dynamics, such as a desired state or unitary evolution, are known a priori.

\subsubsection{Cost function based on black-box access to a target unitary.}

In many NISQ-relevant algorithms, the structure of the quantum circuit need not be prescriptive. Instead, we optimize a general circuit ansatz variationally to minimize some performance cost function. In this setting, we once again may optimize both the discrete and continuous degrees of freedom of the circuit to minimize the cost function. One can conceive of situations where the target circuit unitary $U$ may be given as a black-box process \cite{Chuang1997, Buzek1996}, that is where any decomposition of the circuit is unknown and we do not have a classical description of the unitary entries, or a circuit to compute them. In this situation, the cost function computing the distance between the black-box target $U$ and a given parametric quantum circuit $V(\boldsymbol{\alpha})$ needs to be estimated. 

A possible way to compute distances between a unitary implemented on a quantum computer and a black-box unitary implemented by a different quantum system is the Hilbert-Schmidt test (HST) \cite{Khatri2019}, a generalization of the SWAP test for state preparation, which evaluates the gate fidelity of Eq.~\eqref{eq:gatefid} on a quantum circuit. A related cost function for black-box quantum compilation is given by the local Hilbert-Schmidt (LHS) test -- see Fig.~\ref{fig:rl_ion_diag}, panel (c) --, which has been shown to be easier to optimize and less sensitive to barren plateaus \cite{McClean2018, Cerezo2021}. Observe that the cost function $C_{\text{LHS}}(V, U)$ is bound from above and below by the average gate fidelity given in Eq.~\eqref{eq:gatefid},
\begin{align}
   C_{\text{LHS}}(V, U) < C_{\text{HST}}(V, U) < n C_{\text{LHS}}(V, U),
\end{align}
which implies that minimizing $C_{\text{LHS}}(V,U)$ also minimizes  $C_{\text{HST}}(V, U)$. Other cost functions have been proposed which are based on so-called incoherent learning, i.e., where the quantum computer and the black-box quantum system do not need to interact coherently \cite{jerbi2023power}.

For gradient-based optimization, we need to differentiate the cost functions $C_{\text{HST}}$ or $C_{\text{LHS}}$ with respect to continuous gate parameters.
Unfortunately, the gradient method given in Eq.~\eqref{eq:recursive-grad} requires access to the intermediate unitaries, which are generally not available in real-world scenarios. Moreover, we do not generally have direct access to the gradient of the unitary operations. As a consequence, we need to use the cost function itself to estimate its gradient with respect to the continuous parameters, which proves slower. In fact, computing the gradient in Eq.~\eqref{eq:grad_v_alpha} with the method of finite differences requires $2M$ evaluations of the test circuit, where $M$ is the total number of parameters. However, finite-difference methods applied to quantum circuits tend to have small signal-to-noise ratios and therefore require large numbers of shots \cite{Kyriienko2021}, especially when compared to sampling strategies that estimate the gradient via trigonometric interpolation \cite{Wierichs2022, bittel2022fast}.

Let us now consider the specific case of trapped ions with the gate set introduced in Section \ref{sec:gateset}. There are three different types of gates ($Z$, MS and $C_{xy}$), with four types of parameters shifts: the three rotation angles of the $Z$ gate, the MS gate and the $C_{xy}$ gate, respectively, and the phase term $\phi$ of the MS and $C_{xy}$ gates, which is the same for both unitaries (see also Appendix \ref{appx:parametershift}). \\

In the following, we consider a cost function on the Hilbert Schmidt test, but the results can be applied to the local test as well. In the case of the $Z$ gate, only one parameter $\theta$ is present, whereas for the other two gates we have two possible parameter-shifts. The experimental quantum cost function $C_{\text{HST}}$ comparing the parameterized circuit $V(\boldsymbol{\alpha})$ in Eq.~\eqref{eq:general-circuit} with a target unitary $U$ is given by
\begin{align}\label{eq:quantumcost_2}
    C_{HST}(\boldsymbol{\alpha}) = 1 - \Tr{H (U \otimes V(\boldsymbol{\alpha})^*) \rho  (U^{\dagger} \otimes V(\boldsymbol{\alpha})^T)},
\end{align}
\begin{figure*}[ht!]
	\hspace*{-1cm}
	\centering
	\includegraphics[width = 18cm]{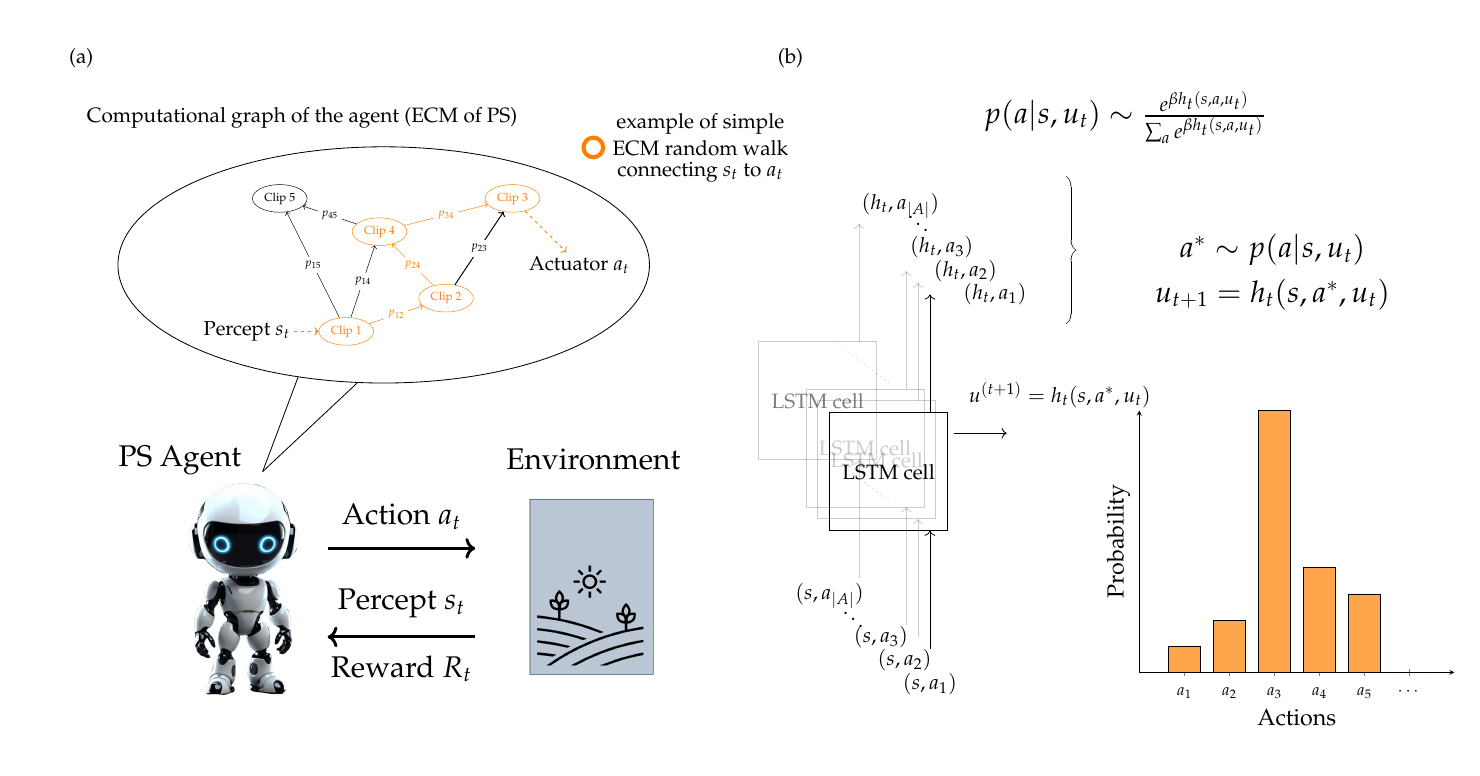}
	\caption{(a) Schematic representation of a PS-environment interaction: the agent is equipped with a memory structure that allows it to process the information input by the environment (in the case of PS, the episodic and compositional memory, ECM \cite{Briegel2012, Melnikov2017}). Every perceptual input from the environment triggers a sequence of transitions  -- showed in orange -- within the internal computational graph of the agent and governed by transition probabilities $p_{ij}$. The sequence of transitions connects a percept clip $s_t$, which in our case corresponds to the sequence of actions $a_1, ..., a_{t-1}$ used to generate the circuit $V_{t}$ using the corresponding gates, to an action clip corresponding to $a_t$. (b) Policy parametrization of the PS-LSTM algorithm. In this implementation, the agent receives a sequential input at time $t$ and constructs a policy by outputting weights $h_t(s,a, u_t)$ corresponding to each action $a$, the current percept $s$ and the hidden state of the LSTM network $u_t$. Afterwards, an action  $a^*$ is sampled from the policy constructed this way and the weight corresponding to this action is propagated further in the hidden state of the LSTM network, as given in Eq.~\eqref{eq:lstm_policy}. The weights are then reset when the episode terminates.}
	\label{fig:pslstm_diagram}
\end{figure*}
where $\rho$ and $H$ are projectors defined in Appendix \ref{appx:parametershift} and Ref.~\cite{Khatri2019}.
We consider here only the shift with respect to one gate at a time, which we name $V_1(\boldsymbol{\alpha}_1)$, while the remaining gates in the circuit, i.e.,  $V_2(\boldsymbol{\alpha}_2), ..., V_L(\boldsymbol{\alpha}_L)$, are fixed. This procedure can be repeated for each gate parameter.
We first consider the case where $V_1(\boldsymbol{\alpha}_1 = \theta) = Z(\theta)$. The exact derivative of $C_{\text{HST}}(\theta)$ is given by:
\begin{align}\label{eq:shift_theta_z}
    \pdv{}{\theta}C_{\text{HST}}(\theta, Z) = C_{\text{HST}}(\theta + \frac{\pi}{2}, Z) - C_{\text{HST}}(\theta - \frac{\pi}{2}, Z)
\end{align}
For $V_1(\boldsymbol{\alpha}_1 = (\theta, \phi)^T) = C_{xy}(\theta, \phi)$ and following Ref.~\cite{Wierichs2022}, the parameter derivative with respect to $\theta$ is given by
\begin{align}\label{eq:shift_theta_xy}
    &\left. \pdv{\theta} C_{\text{HST}}(\theta, \phi, C_{xy}) \right|_{\theta=0} = \\ \nonumber &\sum_{l=1}^{2n} \frac{(-1)^{l-1}}{2 \sin \left(\frac{2l-1}{2n}\pi \right) } C_{\text{HST}} \left(\theta + \frac{2l-1}{2n}\pi, \phi, C_{xy} \right),
\end{align}
whereas for $V_1(\boldsymbol{\alpha}_1 = (\theta, \phi)^T) = \text{MS}(\theta, \phi)$, the parameter-shift rule with respect to $\theta$ becomes,
\begin{align}\label{eq:shift_theta_ms}
    &\left. \pdv{\theta} C_{\text{HST}}(\theta, \phi, \text{MS}) \right|_{\theta=0} = \\ \nonumber &\sum_{l=1}^{2 \lceil \frac{n}{2} \rceil} \frac{(-1)^{l-1}}{2 \sin \left(\frac{2l-1}{\lfloor \frac{n}{2}  + 1 \rfloor}\pi \right) } C_{\text{HST}} \left(\theta + \frac{2l-1}{\lfloor \frac{n}{2} + 1 \rfloor }\pi, \phi, \text{MS} \right) \delta_{i}^{\text{floor}}.
\end{align}
A more simplified rule can be derived for the parameter $\phi$, which is valid for both the $C_{xy}$ and the MS gate

\begin{align}\label{eq:phi_parameter_shift}
    &\pdv{\phi}C_{\text{\tiny HST}}(\theta, \phi) = C_{\text{\tiny HST}}(\theta, \phi + \frac{\pi}{4}) - C_{\text{\tiny HST}}(\theta, \phi - \frac{\pi}{4}) + \\ \nonumber &+ \Big(\frac{1}{\sqrt{2}} - \frac{1}{2}\Big)C_{\text{\tiny HST}}(\theta, \phi - \frac{\pi}{2}) + \Big(\frac{1}{\sqrt{2}} + \frac{1}{2}\Big)C_{\text{\tiny HST}}(\theta, \phi + \frac{\pi}{2}).
\end{align}

Using the expressions given by Eq.~\eqref{eq:shift_theta_z} for the $Z$ gate, by Eq.~\eqref{eq:shift_theta_xy} and Eq.~\eqref{eq:phi_parameter_shift} for the $C_{xy}$ gate and Eq.~\eqref{eq:shift_theta_ms} and Eq.~\eqref{eq:phi_parameter_shift} for the MS gate, we can compute any gradient of cost functions estimated in experiments on quantum devices that use cost functions such as Eq.~\eqref{eq:quantumcost_2}. For more details about the parameter-shift rules of the $C_{xy}$ and MS gates, and further possible simplifications, see Appendix \ref{appx:parametershift}. 

\subsection{Combinatorial Optimization}
\label{sec:rl}

In this section, we consider the problem of determining an optimal arrangement of the gates on the quantum circuit. This problem arises from the necessity of minimizing the depth of a quantum circuit to reduce the total circuit execution time, thereby reducing decoherence. And while the circuit can be compiled in layers \cite{Ostaszewski2021structure}, it is difficult to determine the optimal size due to the large number of possible optimal parameter configurations that produce the same circuit. Therefore, it is necessary to search through various combinations of gate arrangements to determine a minimal one. This task, which partially falls in the realm of combinatorial optimization \cite{Mazyavkina2020}, is particularly suitable for reinforcement learning algorithms \cite{Moro2021}.

\subsubsection{Reinforcement Learning with Projective Simulation}\label{sec:rlps}
Reinforcement learning describes a class of algorithms which use an agent-environment interaction model to maximize a reward function. In particular, the agent can be represented by a parametrized model, called policy, that outputs action signals on the environment upon receiving observations as inputs. For each action or sequence thereof, the environment returns observations and reward signals. In gradient-based methods, the policy parameters are updated in the direction that maximizes the discounted future expected return \cite{Sutton1998}. Based upon the different tasks considered, a vast range of algorithms and methods have been developed to tackle various environments \cite{Li2017}. 

In the following section, we consider the PS architecture -- see Fig.~\ref{fig:pslstm_diagram} (a).

Projective Simulations (PS) is a framework for agency and decision making that has also found applications as a RL agent \cite{Boyajian2020, Melnikov1221}. In that context, a PS agent interacts with an environment by performing actions sampled from an action space $A$, whereas the environment provides the agent with perceptual inputs, which reside in a percept space $S$, and reward signals. Its central feature is represented by a so-called episodic and compositional memory (ECM), see Fig.~\ref{fig:pslstm_diagram} (a), a graphical model consisting of a network, or weighted graph, where the vertices are clips and represent, e.g., remembered percepts or remembered actions, but also more general states of the agent's memory. In this framework, the agent creates a clip inside the ECM each time it receives a previously unknown input from the environment, or each time it creates a new action clip or a more abstract clip, e.g., through action composition. This makes it is possible to create ECM networks with complex graph structures, allowing for generalization capabilities \cite{Melnikov2017, Eva2022}. Each input triggers a random walk between the nodes of the ECM that is governed by the edge weights of the graph. We assume in the following that the ECMs created are two-layered networks, with one layer representing percepts clips and the other action clips and where a percept at time $t$ of the RL interaction is connected directly to all action clips.
The edge weights are initialized uniformly:
\begin{align}
    &\ \forall a \in A, \forall s \in S: h_{0}(s,a) = 1.
\end{align} We define the probability distribution\footnote{In standard PS, the probability is defined without the exponential factors.} over percepts $s \in S$ and actions $a \in A$ as
\begin{align}\label{eq:psh_policy}
    p(a \vert s) = \frac{e^{\beta h_t(s,a)}}{\underset{a}{\sum} e^{\beta h_t(s,a)
    }},
\end{align}
for a edge weight $h_t(s,a)$ connecting $s$ to $a$.
The PS algorithm optimizes the policy, similar to other RL frameworks, through a reward signal, which is provided by the environment to the agent. At each RL time-step $t$, the update rule is given by
\begin{align}\label{eq:h_update}
     h_{t+1}(s, a) = h_t(s, a) - \gamma (h_t(s, a) - 1) + \\ \nonumber + g_t(s,a)R_t,
\end{align}
where $R_t$ is the reward at time step $t$, $\gamma$ is a damping coefficient that regularizes the result and reduces potential instances of trapping in local minima, $g_t(s,a)$ are so-called glow values \cite{Mautner2015}, which are initially set to zero
\begin{align}
    &\ \forall a \in A, \forall s \in S: g_{0}(s, a) = 0.
\end{align}
They are set (or reset) to $g_t(s,a) = 1$ at time $t$ if the corresponding edge is traversed and decay in value for each time step where they are not used according to the rule
\begin{align}
    g_{t+1}(s, a) = (1 - \eta) g_t(s, a),
\end{align}
where $0 \leq \eta \leq 1$. The glow mechanism helps distribute the rewards along the entire chain of state-action transitions traversed by the agent in an episode. We see that the edge weights that are rewarded positively grow, thereby enhancing the probability that the same action is chosen again in the future upon receiving a similar percept.

PS has been successfully extended to include more powerful computational structures, such as deep feedforward energy-based networks \cite{Jerbi2021, Nautrup2020} and recurrent networks \cite{Preti2020}. The aforementioned architectures enable the PS agent to update the policy using function approximators. This allows the agent, as it is the case for deep $Q$-learning \cite{Mnih2015}, to construct more powerful representations of the percept- and action-spaces and achieve high performances in environments with, e.g., continuous parametric percepts and actions without the need of discretization strategies \cite{Sutton1998}. We focus here on the Long-Short Memory Network (LSTM) architecture \cite{hochreiter_thesis, Hochreiter1997}, a recurrent neural network that is equipped with an internal memory architecture that helps it learn long-range correlations in a sequential input. In this case, the action sampled at time $t$ also depends on the LSTM internal state $u_t$, i.e.,
\begin{align}\label{eq:lstm_policy}
    &a^* \sim p(a \vert s, u_t)= \frac{e^{\beta h_t(s, a, u_t)}}{\underset{a}{\sum} e^{\beta h_t(s, a, u_t)}}
\end{align}
with $s \in S$ and $a, a^* \in A$ and the $u$-value is updated w.r.t. the sampled action as follows:
\begin{align}
    u_{t+1} = h_t(s, a^*, u_t). \label{eq:u_update}
\end{align}
The term $u_t = u_t(s,a)$ also depends on percepts and actions, but as we see in Eq.~\eqref{eq:u_update}, only the $u$-value corresponding to the action sampled by the agent at time $t$ are propagated as information to the next RL time step, as shown in Fig.~\ref{fig:pslstm_diagram} (b). This value is the one that carries relevant information about the correlations in the sequential structure of the percepts. We would like to stress that the presence of an additional non-linear term in the update rule in Eq.~\eqref{eq:lstm_policy} induces a different type of ECM structure, where the term $u_t(s,a)$ represents an additional edge weight.
The update of the reward mechanism can be generalized starting from the standard PS reward update mechanism to fit the training of neural network-based policies, in analogy with the case of $Q$-learning \cite{Jerbi2021, Preti2020}.
\section{Circuit compilation}\label{sec:circ_comp}
Quantum compilation is the general task of reproducing a general operation $\mathcal{M} \in \mathbb{C}^{4^n \cross 4^n}$ on a $n$-qubit quantum processor. In particular, we consider the compilation of unitary operations $U \in U(n)$.
\subsubsection{Layer-based compilation and heuristic search}
A general approach for gate synthesis in trapped-ion circuits is discussed in Ref.~\cite{Martinez2016}. This approach is based on the universality of two-qubit entangling gates for quantum computation \cite{DiVincenzo1995}.
As a consequence, it is possible to compile a quantum algorithm in growing layers, i.e., by iteratively placing entangling MS gates on the circuit followed in each case by a collective rotation of the following type
\begin{align}\label{eq:rotation_layer}
    &\mathcal{R}(\theta_k, \phi_k, ..., \theta_{k+n+1}, \phi_{k+1}) = \\ & = C_{xy}(\theta_k, \phi_k) \prod_{i=1}^n Z_i(\theta_{k+i}) C_{xy}(\theta_{k+n+1}, \phi_{k+1}), \nonumber
\end{align}
where the gates $C_{xy}$ and $Z$ are defined in Eqs.~\eqref{eq: ms gate set2}-\eqref{eq: ms gate set3}.
For each layer of gates present on the circuit, we have one MS gate and one general rotation $\mathcal{R}$ -- see Eq.~\eqref{eq:rotation_layer} --, which consists of $n$ local $Z$ gates and two $C_{xy}$ gates.

\begin{algorithm}[H]
	\setstretch{1.20}
	\caption{Exhaustive search with layer-based compilation}
    \label{alg:layercompilation}
	\hspace*{\algorithmicindent} \textbf{Input} Target unitary $U$, threshold $\epsilon$, cost function $C$. \\
	\hspace*{\algorithmicindent} \textbf{Output} $V^*, \bm{\alpha}^*$.
	\begin{algorithmic}[1]
	   \LineComment{Construct layers:}
		\For{$L=1$ to $L_{\text{MS}}$}
		\LineComment{Add MS gate and rotations:}
		\State $V(\boldsymbol{\alpha}) = \prod_{l=1}^L V_l(\boldsymbol{\alpha}_l)$
		\LineComment{All angles but the ones of MS gates:}
		\State $\bm{\alpha} = (\theta_1, \phi_1, ..., \theta_{(n+2)L}, \phi_{2L})^T$ 
		\State $\tilde{K} = (n+2)(L+1)$
		\LineComment{Loop over numbers of rotations:}
		\For{$k=1$ to $\tilde{K}$}
		\LineComment{Loop over combinations of indices:}
		\For{$j=1$ to $\binom{\tilde{K}}{k}$}
		\LineComment{Set angles with chosen indices to zero:}
		\State $\alpha_{\sigma(j,k)} = 0$ 
		\State $\tilde{\bm{\alpha}}^k = (...,\alpha_{\sigma(j,k)} = 0 ..., \alpha_{\sigma(j,k)} = 0, ...)^T$ 
		\LineComment{Cost function according to Eq.~\eqref{eq:rl_cost}:}
		\State $C(\tilde{\bm{\alpha}}^k) = 1 - F(V,U)$ 
		\State $C^* = \min(C(\tilde{\bm{\alpha}}^k), \nabla_{\bm{\alpha}^k} C(\tilde{\bm{\alpha}}^k))$
        \If{$C^* \leq \epsilon$}
        \State $\boldsymbol{\alpha}^* = (\tilde{\bm{\alpha}}^k)^*$
        \State $V^* =  V$
        \State break
        \Else
        \State continue
        \EndIf
		\EndFor
		\EndFor
		\EndFor
	\end{algorithmic}
\end{algorithm}

The unitary added to the circuit at each layer $l$ is:
\begin{align}
    V_l(\boldsymbol{\alpha_l}) = \text{MS}(\theta^{\text{MS}}_{l}, \phi^{\text{MS}}_{l}) \ \mathcal{R}(\theta_l, \phi_l, ..., \theta_{l+n+1}, \phi_{l+1}).
\end{align}

\begin{algorithm}[H]
	\setstretch{1.20}
	\caption{PS-based compilation}
	\label{alg:ps_comp}
	\hspace*{\algorithmicindent} \textbf{Input} Target unitary $U$,  Action set $G_1 = \text{MS}, G_2=C_{xy}, G_3=Z_1, ..., G_{n+2}=Z_n$, threshold function $\epsilon_t$, cost function $C$ \\
	\hspace*{\algorithmicindent} \textbf{Output} $V^*, \bm{\alpha}^*$
	\begin{algorithmic}[1]
		\For{$e=1$ to $E_{\text{max}}$}
		\State $s_1=(0, ..., 0), V_1 = I_d$
		\For{$t=1$ to $L_{\text{max}}$}
		\LineComment{Sample action according to Eq.~\eqref{eq:psh_policy}:}
		\State $a_t \sim \pi(a \vert s_t)$
		\State $V_{t+1}(\bm{\alpha}_t) = G_{a_t} V_{t}(\bm{\alpha}_{t-1})$
		\State $s_{t+1} = (a_{1}, ..., a_t, 0, ..., 0)$
		\State $\boldsymbol{\alpha} = (\boldsymbol{\alpha}_1, ..., \boldsymbol{\alpha}_t) \sim 2\pi \cdot \mathcal{N}(\boldsymbol{0}_{\boldsymbol{\alpha}}, \boldsymbol{I}_{\boldsymbol{\alpha}})$ 
		\LineComment{According to Eq.~\eqref{eq:rl_cost}:}
		\State $\boldsymbol{\alpha}^* = \text{argmin}(C(\bm{\alpha}), \nabla_{\boldsymbol{\alpha}} C(\bm{\alpha}))$  
		\State $R_t = \begin{cases}
		2 & \text{if $\epsilon_{\text{min}} \leq C(\boldsymbol{\alpha}^*) \leq \epsilon_t$}\\
		10 & \text{if $C(\boldsymbol{\alpha}^*) \leq \epsilon_{\text{min}}$}\\
		0 & \text{otherwise}	    
		\end{cases}$
		\If{$C^* \leq \epsilon_{t}$}
		\State $V_* = V_{t+1}$
		\State break
		\EndIf
		\State Update rule for all $h_{t+1}(s_t, a_t)$ (see Eq.~\eqref{eq:h_update})
		\EndFor
		\State Update threshold $\epsilon_t$ according to Eq.~\eqref{eq:threshold_update}
		\EndFor
	\end{algorithmic}
\end{algorithm}

The cost function
\begin{align}\label{eq:rl_cost}
    C(\bm{\alpha}) = 1 - \frac{1}{d^2} \abs{\tr{\prod_{l=1}^L V(\bm{\alpha}_l) U^{\dagger}}}^2,
\end{align}
based on the gate fidelity in Eq.~\eqref{eq:gatefid}, has to be minimized with respect to the angle parameters $\boldsymbol{\alpha}$. After obtaining the optimal parameters $\boldsymbol{\alpha}^*$, if the desired error threshold is reached, the algorithm terminates, otherwise a new layer of gates is placed on the circuit, up to a maximum number of layers $L_{\text{MS}}$. By running this procedure iteratively with different angle parameter sets, we can search through the optimization landscape to find different gate decompositions (see Algorithm~\ref{alg:layercompilation} and Ref.~\cite{Martinez2016}).

However, while this method can be applied to circuits with a small number of layers and qubits, its use becomes impractical as these two parameters grow. In fact, the number of total combinations to analyze for a given number of MS layers $L_{\text{MS}}$ and a number of qubits $n$ is
\begin{align}
    N_{\text{combinations}} = \frac{2^{(n+2)(L_{\text{MS}} + 1)} - 1}{1 - \frac{1}{4}^{ (n+2)}} - L_{\text{MS}}^2.
\end{align}
For a $3$-qubit circuit with $L$ layers of entangling gates, the number of possible circuits is $N_{\text{combinations}}=1049591$. For a $4$-qubit circuit with $5$ entangling gates the number of combinations is approximately $N_{\text{combinations}} \sim 2^{36}$. We see that already for $4$-qubit gates, a brute force approach is unfeasible. Moreover, due to the large search space, even approaches based on random search are not a viable option, especially if the number of entangling gates is large. \\
In the following section, we suggest a different approach to the optimization scheme where the position of the discrete parameters is modified by a RL agent equipped with a curriculum scheme, whereas the continuous parameters are optimized at each iteration and using a pre-defined heuristic for angle initialization. This can offer benefits in several situations where the number of gates is large enough to make the use of automated search prohibitive but not so large to make the problem completely intractable from the point of view of continuous parameter optimization.

\subsubsection{Reinforcement learning-based compilation}
In the following, we discuss the implementation of RL-based compilation. In this system, the agent acts on the environment, which represents a quantum circuit, by choosing a gate from, e.g., the gate set of Section \ref{sec:gateset} and placing it on the circuit. As an observation, the agent receives information about the environment internal state. This can vary depending on the task to be solved: in Ref.~\cite{Moro2021}, e.g., the input is a single-qubit unitary and the task is to construct a pre-trained optimal compiler that constructs any unitary with minimal average number of actions. In Refs.~\cite{Preti2020, Bukov2018}, the agent receives an encoded representation of the quantum circuit in terms of gates and qubits. The circuit-based input has the advantage of scaling linearly with the number of qubits for $n$-qubit entangling gates and its sequential structure makes it suitable for recurrent or autoregressive policies \cite{yao2021reinforcement, Sivak2022}. 

The perceptual input of the PS-LSTM agent is given by the current gate on the circuit. The action of the PS-LSTM agent is placing one further gate on the circuit. A representation of the interaction between the agent and the quantum circuit environment is given in Fig.~\ref{fig:rl_ion_diag}: at each time step, the agent -- Fig.~\ref{fig:rl_ion_diag} (a) -- can place one or more gates on the circuit. Then the circuit -- Fig.~\ref{fig:rl_ion_diag} (b) -- is mapped to the corresponding unitary function, which depends on parameters $\boldsymbol{\alpha}$. The gradient-based optimizer -- Fig.~\ref{fig:rl_ion_diag} (d) --, i.e., the L-BFGS-B algorithm \cite{Liu1989}, is given the task of finding the optimal set of parameters to maximize the fidelity -- Fig.~\ref{fig:rl_ion_diag} (c) -- between the current circuit and a target unitary process. The reward function and the percept for the next interaction step are constructed based on the result of the optimization. The algorithm is shown in Algorithm~\ref{alg:ps_comp} for the standard PS method and can be easily generalized to a framework with deep networks \cite{Jerbi2021, Preti2020}.

Furthermore, one may design different types of RL environments based on the way the agent places gates on the circuit. We develop here two different approaches for two different types of RL environments: In the first environment, we just mimic the structure of layer-based compilation, in which we keep the entangling gates fixed and then let the agent vary the rotations between them. This assumes that the RL interaction is defined by the number of gates placed between each entangling layer and the total amount of entangling layers considered. This architecture has the advantage of restricting the search space of the RL agent, but it allows for less exploration of the cost function landscape.

\begin{figure*}[ht!]
    \hspace{-0.5cm}
    \includegraphics[width=\textwidth]{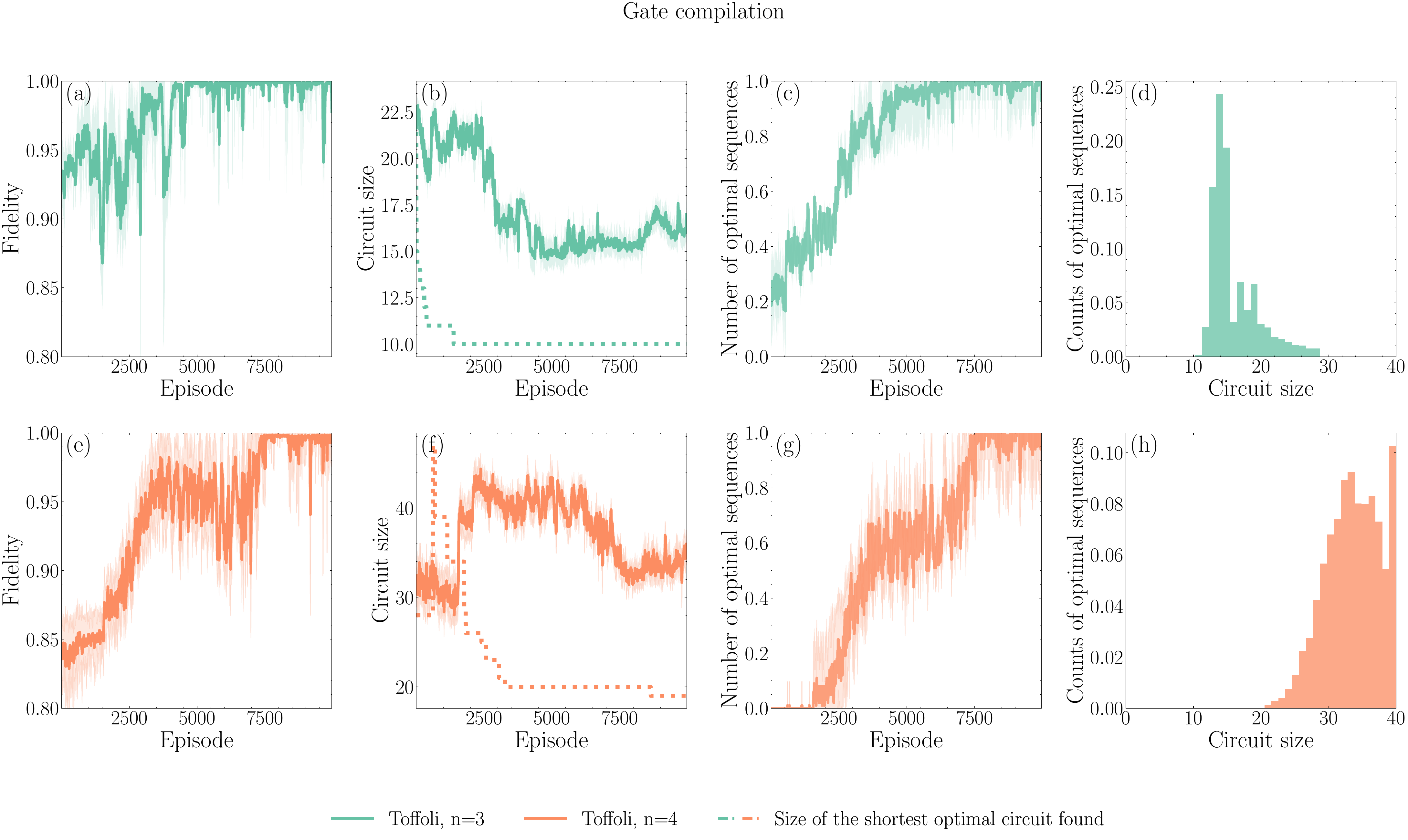}
    \caption{Quantum circuit optimization of different gates using the gate set defined in Eqs.~\eqref{eq: ms gate set}-\eqref{eq: ms gate set3}. (a), (e) show the average reward; (b), (f) the average circuit size (thick line) and the size of the best circuit found so far by all agents (dotted line); (c), (g) the number of optimal sequences per episode; (d), (h) are histograms which show the distribution of the optimal gate sequences over the circuit size. All lines refer to simulations of the $3$-qubit Toffoli gate, first compiled on a $3$-qubit (green line, average over 10 agents and 20 episodes) and then on a $4$-qubit circuit (orange line, average over 5 agents and 20 episodes). Due to the $n$-qubit interaction of the trapped-ion gates, the optimal sequence that generates the gate -- which we define as the shortest sequence whose infidelity falls below $\epsilon_{\text{min}} = 10^{-2}$ --  increases in size from $10$ to $19$ gates. Shaded regions in the plots represent the corresponding standard deviations. While the average fidelity increases to reach the maximum for both circuits, we see that the average circuit length appears to be higher than the shortest circuit length. There are possible explanations for this: the shortest sequences can be harder to optimize, so there is a higher chance that the optimizer fails at outputting the cost function minimum for a given circuit structure and the curriculum scheme, that modifies the problem online during the training. Overall, we observe that the size of the optimal circuit decreases as training progresses, thus validating the successful optimization of the policy.}
    \label{fig:gate_compilation}
\end{figure*}
The second architecture allows the agent to place any available gate, entangling or non-entangling, on the circuit and as such does not restrict the action of the agent on the quantum circuit, with one single exception: if the agent places the same type of gate twice in a row on the circuit, these two gates are merged together in one single gate. This is needed to avoid the agent getting stuck in a loop where it keeps performing the same operation over and over again, without any meaningful exploration of the optimization landscape from the perspective of the continuous optimizer. We employ this architecture in our simulations.

In our implementation, the action $a_t$ chosen at time $t$ is the index of certain gate in the gate set, whereas the corresponding percept $s_t$ is given by the concatenation of previously chosen actions $(a_1, ..., a_{t-1})$. Thus, the action space is given by the gate indices $A = \{ 1, 2, ..., n + 2 \}$  and the percept space by the Cartesian product of $L_{\text{max}}$ action spaces: $S = A^{\times L_{\text{max}}}$. The PS-LSTM algorithm, however, can also be given just the action at time $t-1$ as percept, since the LSTM memory can automatically capture the correlations between the elements in the sequence. Here, we adopt a RL training procedure with a curriculum scheme as described in Ref.~\cite{Ostaszewski2021}. This allows the agent to sufficiently explore the solution space and gradually adapt the solution. More specifically, for each task of quantum circuit optimization, we define a curriculum strategy where we reward the agent at time step $t$ each time it finds a sequence with achieved minimal infidelity $C(\boldsymbol{\alpha}^*)$ lower than a chosen moving threshold $\epsilon_t$ and a fixed threshold $\epsilon_{\text{min}} = 10^{-2}$:
\begin{align}\label{eq:reward}
    R_t = \begin{cases}
      2 & \text{if $\epsilon_{\text{min}} < C(\boldsymbol{\alpha}^*) < \epsilon_t$}\\
      10 & \text{if $C(\boldsymbol{\alpha}^*) < \epsilon_{\text{min}}$}\\
      0 & \text{otherwise}
    \end{cases}.    
\end{align}
The episode terminates when the reward the cost function minimum in a given time step falls below the threshold $\epsilon_t$ or when the maximal length of the circuit per episode, $L_{\text{max}}$, is reached. The RL training terminates upon reaching the maximum number of episodes $E_{\text{max}}$. The reward scheme helps to progressively increase the fidelity throughout training without allowing for too long circuits.
The threshold is then lowered as episodes progress based on previous rewards obtained by the agent. In our implementation, we lower the threshold when it has been surpassed by the agent at least 500 times using the following scheme:
\begin{align}\label{eq:threshold_update}
    \epsilon_{t+1} = \begin{cases}
      \epsilon_{\text{min}} + \frac{1}{2}(\epsilon_{\text{min}} - \epsilon_t) & \text{if $\epsilon_{\text{min}} \leq \epsilon_t \leq 1$}\\
      \epsilon_{\text{min}} & \text{if $\epsilon_t \leq \epsilon_{\text{min}}$}\\
      1 & \text{otherwise}.
    \end{cases}    
\end{align}

\section{Results}\label{sec:results}
We test our algorithm on two relevant tasks of circuit optimization: Standard gate compilation, which can be useful in particular for experimental applications of frequently used gates (Toffoli, etc.) and the simulation of black-box unitary processes with quantum circuits. The latter framework is particularly interesting for quantum simulation and offers the possibility of implementing our algorithm directly on an experimental setting where a black-box unitary process has to be simulated by a quantum circuit with available gates.

\begin{figure*}
    \hspace{-0.5cm}
    \includegraphics[width=\textwidth]{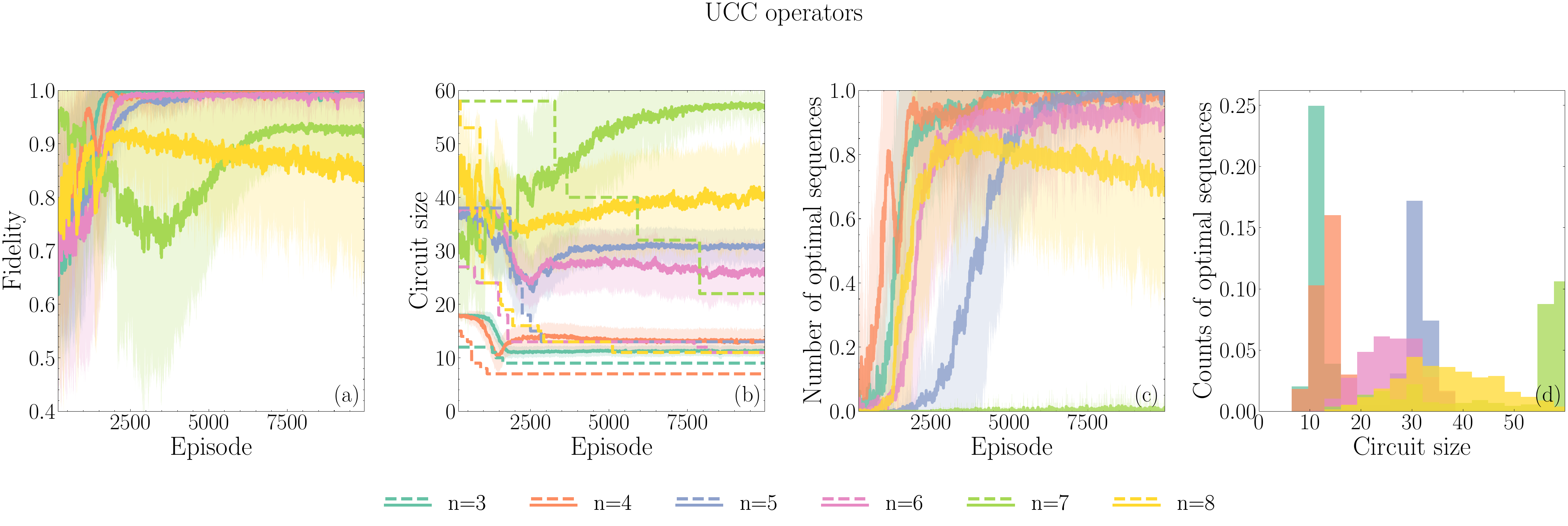}
    \caption{(a) Fidelity; (b) circuit size (thick lines) and size of the best circuit found by all agents (dotted lines); (c) number of optimal sequences per episode and (d) histogram of the optimal sequences -- i.e., whose infidelity falls below $\epsilon_{\text{min}} = 10^{-2}$ -- based on the circuit size for the UCC operators defined in Eq.~\eqref{eq:ucc_def} for $\beta = \frac{\pi}{2}$ and for 3, 4, 5, 6, 7, 8 qubits. The learning curves show the average over 20 agents, after which an average over a time window of 20 episodes is performed. We observe how the average maximal fidelity reached by the agents decreases as a function of the number of qubits, whereas the optimal circuit size increases, which means that a longer circuit is necessary to synthesize the desired operator. Moreover, the fraction of optimal sequences found by the agent is also decreasing for higher numbers of qubits, as optimal policies become more and more sparse and thus harder to discover for the PS-LSTM agent. The hardest task for the PS-LSTM agent proves to be $7$-qubit UCC operator, where the learning curve of the agents fails to converge, while, e.g., for the $8$-qubit UCC unitary the agent can find sequences with very high fidelities ($1 - F < 10^{-5}$), even though the convergence is sub-optimal. We also observe that for each learning curve there is a dip at some point in the training: this is most likely due to the curriculum scheme, which modifies the reward threshold during training and therefore influences the size of the optimal circuits found by the agent. However, when considering the ensemble of simulations, we see that the agents are capable of finding shorter and shorter optimal circuits. Shaded regions in the plots represent the corresponding standard deviations.}
    \label{fig:ucc_operators}
\end{figure*}

\subsection{Example 1: gate compilation}
As a first application of the method presented above, we consider the problem of compiling a gate on an ion-trap quantum processor using the gate set given in Eqs.~\eqref{eq: ms gate set}-\eqref{eq: ms gate set3}.
For this class of tasks, we choose two gate compilation problems which can be of interest for typical applications, before passing to a more general framework which considers black-box unitaries.
First, we consider a standard quantum computing gate, the $3$-qubit Toffoli gate \cite{Toffoli1980, Monz2009} -- see Fig.~\ref{fig:gate_compilation} -- for a $3$-qubit (green line, upper four plots) and a $4$-qubit circuit (orange line, lower four plots).
The dashed lines in Fig.~\ref{fig:gate_compilation} show the optimal solutions found by the agent. In the $3$-qubit case, this solution matches agrees with the results given in Ref.~\cite{Martinez2016}.
Fig.~\ref{fig:gate_compilation} (a), (e) show the fidelity of the sequences produced and optimized by the agents as the learning progresses, (b), (f) show the average circuit size and the size of the shortest circuit found by the agent, (c), (g) show the average number of optimal sequences, i.e., with fidelity higher than 0.99, found in each episode (which in the best-case scenario should converge to one per episode) and (d), (h) show histograms representing the number of sequences for different bins of circuit size. We observe from the first and second plot in each row of Fig.~\ref{fig:gate_compilation} that the agents are capable of maximizing the fidelity and at the same time of reducing the average circuit size in both cases. Moreover, while the average circuit size is larger than the size of the best circuit, most likely due to the presence of the curriculum and the influence of local minima on the angle optimization, we also see that the agents find shorter and shorter optimal circuits as the training progresses, hinting that the at least one agent in the ensemble is indeed learning to optimize the circuit correctly. Moreover, the third plot in each row shows us that the agents find an increasing number of high-fidelity sequences during training. The minimal sequences found by the agents are located in the leftmost tail of the distribution. 
The $3$-qubit Toffoli gate implemented on a $4$-qubit circuit could have in principle a relatively long generating quantum circuit in the chosen gate set, since the 4-qubit gate set also affects the qubits left unchanged by the Toffoli gate. We also observe, in our simulations, that sequences generating this gate are sparse in the action space, which could make it generally difficult to produce this gate on a register of $n$ qubits without the help of MS and $C_{xy}$ gates acting only on subspaces of the register. We test whether our method can discover a circuit of reasonable size. 
For large numbers of qubits and large circuit sizes, we observe that although the algorithm can find shorter circuits, it is still impaired by the computational overhead of simulating such circuits exactly. In this case, the gate decomposition method introduced in Section \ref{sec:dynamics} provides a useful speedup for RL simulations, especially if compiled on GPU architectures.

\subsection{Example 2: General quantum process simulation of a Hamiltonian model}
As a second example, we compile parametric-type operators that play a relevant role in quantum chemistry \cite{Taube2006, Bartlett2007}, i.e., UCC-type operators. These are part of a more general class of many-body operators \cite{Motzoi2017}:

\begin{figure*}[ht!]
	\hspace{-0.5cm}
	\includegraphics[width=\textwidth]{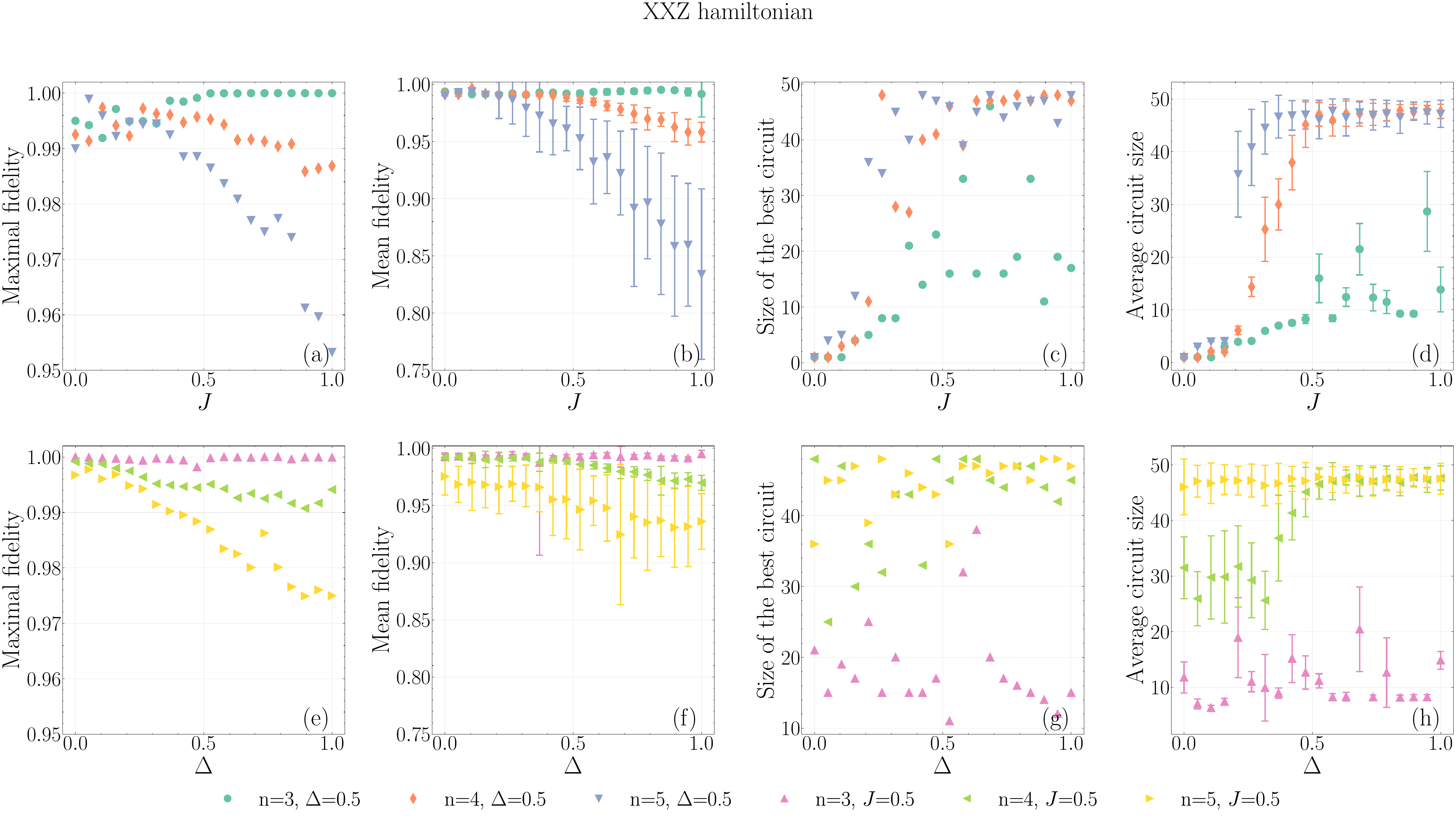}
	\caption{These plots summarize the result of simulating the XXZ-model unitary with our approach for different parameter configurations for $n=3, 4, 5$ qubits (3 agents for each parameter sample). (a) shows the maximal value of the fidelity found by the agents, (b) the mean fidelity, (c) the size of the shortest circuit associated with the fidelity in (a) and (d) the average circuit size, as a function of the coupling $J$ of the XXZ model for two different configuration of the transverse coupling $\Delta$ = 0.5 with varying $J$ and and $J$ = 0.5 with varying $\Delta$ (dark green, orange and blue and pink, light green and yellow, respectively). The evolution time was fixed to $\tau=0.25$ and the maximum size of the circuit to 50. The average fidelity is calculated over the last 200 episodes. Vertical bars show the standard error for the mean values of fidelity and circuit size. We see here that the circuit size increases dramatically as the parameters $\Delta$ and $J$ increase. We also see from the number of outliers in (c) and (g), that for certain values of the parameters it is hard for the agents to exactly retrieve the optimal circuit.}
	\label{fig:xxz_analysis}
\end{figure*}

\begin{align}\label{eq:ucc_def}
    U(\beta) = \exp(i \beta \left(\prod_{i=1}^n (\sigma^{(i)}_x - i\sigma^{(i)}_y) + \text{h.c.} \right)).
\end{align}

The class of operators defined by Eq.~\eqref{eq:ucc_def} resembles the collective rotations used in trapped-ion gate sets given in Eqs.~\eqref{eq: ms gate set}-\eqref{eq: ms gate set3}, they are however sparser.
Due to their importance, they represent our first candidate for process simulation on the quantum circuit. The results for the simulation using PS-LSTM of several such operators -- $n=3,4,5,6, 7, 8$ -- and for $\beta=\frac{\pi}{2}$ is given in Fig.~\ref{fig:ucc_operators}. We observe that the agent is able to increase the fidelity up to the optimal value.  We also see that the size of the optimal circuit generating the corresponding operators increases with the number of qubits.  Furthermore, due to the growing number of local rotations available to the agent and the sparseness of the reward, it becomes increasingly difficult to find and reward optimal sequences. This also shows the benefit of implementing curriculum strategies in such hybrid discrete-continuous optimization problems, where the landscape both for the RL agent and for the numeric optimizer become increasingly sparse. We also note that the generation of $7$ and $8$-qubit UCC operators proves more challenging.  In particular, for the $7$-qubit UCC unitary, the agent is able to raise the fidelity to values close to $F=0.99$, but it cannot significantly increase the number of circuits with $F > 0.99$ over the course of the training. In the case of the $8$-qubit UCC operator, the agent instead proves capable of discovering sequences with very high fidelity, i.e $1-F < 10^{-5}$, although the average fidelity worsens slightly towards the end of the training. We also see that an ensemble of agents can successfully reduce the size of the best circuit, even in those cases where the average performance worsens during training.

We would like now to consider a (black-box) unitary $U(\tau) = e^{-iH \tau}$, where $\tau$ is the evolution time described by a Hamiltonian of the following type:

\begin{align}\label{eq:ham_simple}
    &H = -2h \sum_{i = 1}^n \sigma^{(i)}_z - J \sum_{i=1}^{n-1} \Bigg(  \sigma^{(i)}_x \otimes  \sigma^{(i+1)}_x + \sigma^{(i)}_y \otimes  \sigma^{(i+1)}_y \\ + \nonumber
    &\Delta \left(\sigma^{(i)}_z \otimes  \sigma^{(i+1)}_z - \frac{1}{4}I \right) \Bigg),
\end{align}
which is usually referred to as the XXZ model \cite{takahashi_1999}. 
We want to analyze how the optimal quantum circuit found by the RL agent changes when we vary the Hamiltonian parameters. For the XXZ model, this parameter variation represents for instance the transition from a XX model when $\Delta=0$ to a XXX model for $\Delta=1$, or from $ZZ$ interaction for $J=0$ to a XXZ model for $J>0$. In fact, we observe in different parameter regions different behaviours of the circuit structure.
Results of several RL runs  for two different configurations, one with fix coupling $J=0.5$ and varying $\Delta$ and another one with fix $\Delta=0.5$ and varying $J$ of the XXZ model unitary are shown in Fig.~\ref{fig:xxz_analysis} for 3, 4 and 5 qubits. Plots (a), (e) show the maximal fidelity found by the agents for each run, plots (b), (f) the mean fidelity, plots (c), (g) the number of gates in the shortest circuit among those with the highest fidelity and plots (d), (h) the average circuit size, all represented in their functional dependence from the coupling $J$ and transverse coupling $\Delta$ of the XXZ model for two different configuration of the transverse coupling $\Delta$ = 0.5 with varying $J$ and and $J$ = 0.5 with varying $\Delta$ (dark green, orange and blue and pink, light green and yellow, respectively). We also see how rapidly the circuit size grows in the presence of entanglement, for example from $J=0$ to $J=1$, whereas the increase in average circuit size seems less pronounced as we vary $\Delta$. We observe that the optimal circuit size can also decrease, for example when $\Delta=1$ and $J=0.5$. We also see some instabilities and high variance in both the average circuit size and the size of the best circuit discovered, though it is hard to determine whether the agent fails at finding a shorter circuit for a specific parameter or the problem becomes suddenly harder to represent with the given gate set due to the parameter variation.
\section{Conclusion}
In this work we construct a framework to learn both classically simulated and black-box unitaries on a quantum circuit using RL and unconstrained optimization. Our simulation is specifically tailored to an ion-trap architecture based on collective gates and local rotations. We demonstrate the synthesis of optimal circuits for Toffoli gates and UCC operators for varying numbers of qubits. As instances of black-box unitaries, we also consider Hamiltonian simulations of the XXZ model. More specifically, we study the convergence and the quality of the solutions found by agent and optimizer as we modify relevant parameters of the underlying black-box unitary, such as the coupling in the XXZ Hamiltonian.
After testing the algorithm on different unitary process simulation scenarios, we observe that the optimization of circuits is generally possible even for large numbers of gates, it is however difficult to foresee how sparse the cost function minima can be as we increase the number of qubits and reduce the sparsity in the corresponding Hamiltonian. Possible improvements include combining the RL search with a graph traversal algorithm to have a more efficient exploration of the discrete action space \cite{dalgaard2020global, Zhang2020}. We expect that this approach can be applied to different architectures beyond trapped ions, and more carefully engineered reward functions can be developed to further enhance the discovery of optimal circuits. Our unified optimization, combining circuit compilation and unitary synthesis, may find use both in classical pre-optimization and in experimental circuit learning, be it for in-situ (variational) algorithms or module-compilation tasks.  

\section{Data and Code availability}
The code and the data are available at the following links: \href{https://github.com/franz3105/RL_Ion_gates}{\url{https://github.com/franz3105/RL_Ion_gates}} and \href{https://zenodo.org/records/8288977}{https://zenodo.org/records/8288977}.

\section*{Acknowledgement}
We would like to thank Markus Schmitt, Robert Zeier and Marco Canteri for the useful discussions. FP, MS and FM acknowledge support from the German Federal Ministry of Education and Research (BMBF) within the framework programme "Quantum technologies – from basic research to market" (Project QSolid, Grant No. 13N161), the European Union's Horizon Programme (HORIZON-CL4-2021-DIGITALEMERGING-02-10) Grant Agreement 101080085 QCFD and HORIZON-CL4-2022-QUANTUM-02-SGA via the project 101113690 (PASQuanS2.1), by the Deutsche Forschungsgemeinschaft (DFG, German Research Foundation) under Germany’s Excellence Strategy – Cluster of Excellence Matter and Light for Quantum Computing (ML4Q) EXC 2004/1 – 390534, and from the Jülich Supercomputing Center through the JUWELS and JURECA cluster. SJ acknowledges the Austrian Academy of Sciences as a recipient of the DOC Fellowship.
This research was funded in whole or in part by the Austrian Science Fund (FWF) through the DK-ALM W$1259$-N$27$ ([Grant DOI: 10.55776/W1259] and the SFB BeyondC F7102 [Grant DOI: 10.55776/F71]. For open access purposes, the authors have applied a CC BY public copyright license to any author-accepted manuscript version arising from this submission. This work was also co-funded by the European Union (ERC, QuantAI, Project No. $101055129$). Views and opinions expressed are however those of the author(s) only and do not necessarily reflect those of the European Union or the European Research Council. Neither the European Union nor the granting authority can be held responsible for them. Simulations were realized using the libraries NUMBA \cite{lam2015numba}, JAX  \cite{jax2018github}, PYTORCH \cite{NEURIPS2019_9015} and SCIPY \cite{2020SciPy-NMeth}. The robot picture in Fig.~\ref{fig:rl_ion_diag} and Fig.~\ref{fig:pslstm_diagram} was generated by the authors using Midjourney.

%\newpage

%

\newpage
\onecolumngrid
\appendix
\section{Fast analytic ion gates}\label{sec:fast_gates}
\begin{figure}
\centering
\includegraphics[width=\textwidth]{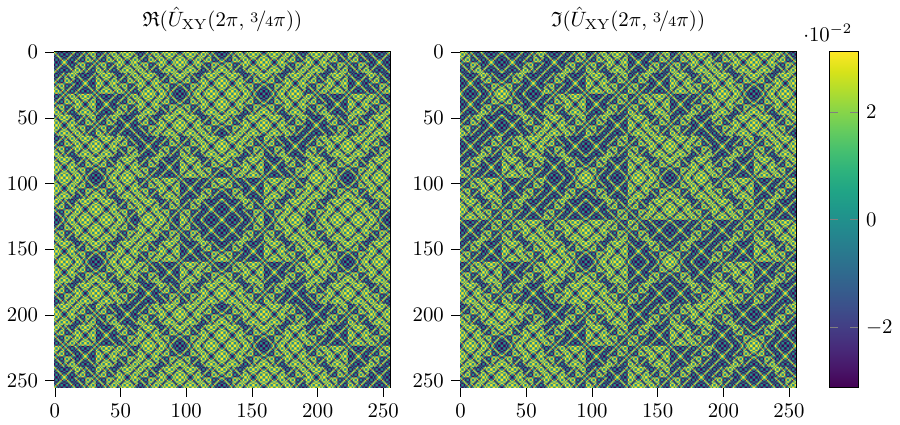}
\caption{ Real and imaginary components of the unitary matrix of the $n$-qubit XY-rotation gate with $\phi=2\pi$ and $\theta=\nicefrac{3}{4}\pi$. The colormap describes the value range of the matrix elements. \label{fig: real_imag_u_xy}}
\end{figure}

In this section, we derive the representations for the MS and $C_{xy}$ gates based on their spectral decomposition and we show how these can be further simplified. We also provide descriptions of the gate gradients based on the optimized representations. The $n$-qubit XY-Rotation and MS Hamiltonians are given by
\begin{align}
\hat{H}_\text{XY}(n, \phi) =&  S_x \cos(\phi) + S_y  \sin(\phi),\\
\hat{H}_\text{MS} (n, \phi) =& \hat{H}_\text{XY}(n, \phi)^2 = \left(  S_x \cos(\phi) + S_y \sin(\phi) \right)^2,
\end{align}
We will first focus on solving the XY-rotation gate and then generalize our solution to the MS gate, using the fact that the MS-Hamiltonian is the square of the XY-Hamiltonian. A representation of the $XY$ unitary in terms of its real and imaginary entries for $n=8$, $\theta=\frac{3}{4}\pi$ and $\phi=2\pi$ is given in Fig.~\ref{fig: real_imag_u_xy}.  
\subsection{General Approach - Constructing the XY-rotation Gate}

Due to the lack of multi-qubit interaction terms in the XY-rotation Hamiltonian, the eigenvectors and eigenvalues of the $n$-qubit case can be constructed from the single qubit closed-form solution, allowing us to factorize the evolution into $\phi$ dependent terms and $\theta$ dependent terms ($\theta$ can be understood as the time evolution parameter, using a standard notation of the literature of trapped-ion quantum computing). This allows us to calculate the associated unitaries from element-wise operations. We decompose the Hamiltonian into
\begin{align}
\hat{H}_\text{XY}(n, \phi) = \hat{V}_n(\phi)  \Lambda_\text{XY} \hat{V}_n^\dagger(\phi), \label{eq:xy_diagonalisation} 
\end{align}
with the $\phi$ dependent $\hat{V}_n$ and the diagonal matrix of the eigenvalues $\Lambda_\text{XY}$. For the unitaries the eigenvalues then capture the $\theta$ dependence via $\exp(-i\Lambda_\text{XY} \frac{\theta}{2})$. 

\paragraph{Eigenvalues} The eigenvalues are constructed from the single qubit case using the following Kronecker sum notation $\hat{A}^{\oplus n}= \sum_{i=1}^n \mathds{I}^{\otimes i-1} \otimes \hat{A} \otimes \mathds{I}^{\otimes n-i}$,
 \begin{align}
 \Lambda_\text{XY} = \text{diag}\left(2 \mathbf{b}_n - n\right), \quad \text{with} \quad  \mathbf{b}_n = \begin{pmatrix} 0 \\ 1 \end{pmatrix}^{\oplus n}.
\end{align}

We find that the $i$-th eigenvalue can be constructed from the binary Hamming weight vector $\mathbf{b}_n$, because the $i$-th component of the binary Hamming weight $b_n(i)$ corresponds to the number of qubits with $\ket{1}$ at index $i$ of the Hilbert space.

\paragraph{Eigenvectors}
The eigenvector matrices $\hat{V}_n(\phi)$  are also constructed from the single qubit case, but using the tensorproduct, 
\begin{align}
 \hat{V}_n(\phi) = \frac{1}{\sqrt{2^n}}\begin{pmatrix} 1 & 1 \\ -e^{i\phi} & e^{i\phi} \end{pmatrix}^{\otimes n} = \hat{P}_n(\phi) \odot  \frac{1}{\sqrt{2^n}}\underbrace{(\hat{\mathds{I}}_2 +i \sigma_y)^{\otimes n}}_{= \hat{S}_n}. \label{eq:xy_decomposition_matrices}
 \end{align}
We furthermore decompose them into an elementwise product ($\odot$) of a sign $\hat{S}_n$ and phase $\hat{P}_n(\phi)$ component. 
The phase components $\hat{P}_n(\phi)$ are column-independent $\hat{P}_n(\phi) = [ \mathbf{p}_n (\phi), \mathbf{p}_n (\phi), \cdots, \mathbf{p}_n (\phi)]$, where the vectors themselves can be regarded as the element wise complex exponential $v_\phi$ of the binary Hamming weight vector $\mathbf{b}_n$
\begin{align}
\mathbf{p}_n(\phi) = \begin{pmatrix} 1 \\ e^{i \phi} \end{pmatrix}^{\otimes n} = \exp(i\phi \mathbf{b}_n) = v_\phi(\mathbf{b}_n).
\end{align}

\paragraph{Calculating the Unitary}
We can now use the column independence of the phase components $\hat{P}_n(\phi)$, to commute it with the eigenvalues. This allows us to separate the phase component further. For the unitary of the XY-gate we then find
\begin{align}
C_{xy}(\phi, \theta) = & \exp\left( -i \hat{H}_\text{XY}(\phi) \theta \right) = \hat{V}_n(\phi) \exp(-i \Lambda_\text{XY})\theta) \hat{V}_n^\dagger(\phi) \nonumber\\
=& \frac{1}{2^n} \underbrace{\hat{P}_n(\phi)\hat{P}_n^\dagger(\phi)}_{=v_\phi(\hat{C})} \hat{S}_n \exp(-\nicefrac{i\theta}{2}\Lambda_\text{XY}) \hat{S}_n^\dagger.
\end{align}
We substitute the phase term product with the previously defined element wise complex exponential of $\hat{C} = \hat{P}_n - \hat{P}_n^T$. 
Furthermore we split the sign components into degeneracy subspaces one for each different eigenvalue in $\lambda_\text{XY}$. We decompose $\hat{S}_n = \sum_{i=0}^n \hat{S}_{\lambda_i}$, where $\hat{S}_{\lambda_i}$ retains the values of $\hat{S}_n$ for columns with binary Hamming weight $i$, but is zero for all other columns. From this we can then construct cached matrices $\hat{D}_{\lambda_i} = \hat{S}_{\lambda_i} \hat{S}_{\lambda_i}^\dagger$, so that we can rewrite
\begin{align}
    \hat{S}_n \exp(-\nicefrac{i\theta}{2}\Lambda_\text{XY}) \hat{S}_n^\dagger =\sum_{i=0}^n e^{-i\lambda_{\text{XY},i} \frac{\theta}{2}} \hat{D}_{\lambda_i}.
\end{align}
 From this we conclude
\begin{align}
C_{xy}(\phi, \theta) = & \frac{1}{2^n} v_\phi(\hat{C}) \odot \sum_{i = 0}^n \exp(-i \lambda_\text{XY,i} \theta) \phantom{,} \hat{D}_{\lambda_i}.\label{eq:xy_exp}
\end{align}
The element wise exponential $v_\phi(\hat{C})$ is constructed efficiently by reusing $2n+1$ phase values, because $C_{ij} \in [0,1,\cdots, 2n]$. Each of the required computations require $\mathcal{O}\left((2^n)^2\right)$ operations, resulting in a total complexity of $\mathcal{O}\left((n+1)(2^n)^2\right)$. 

\subsection{Generalization to the MS gate}
\begin{figure}
\hspace{-0.5cm}
\centering
\includegraphics[width=\textwidth]{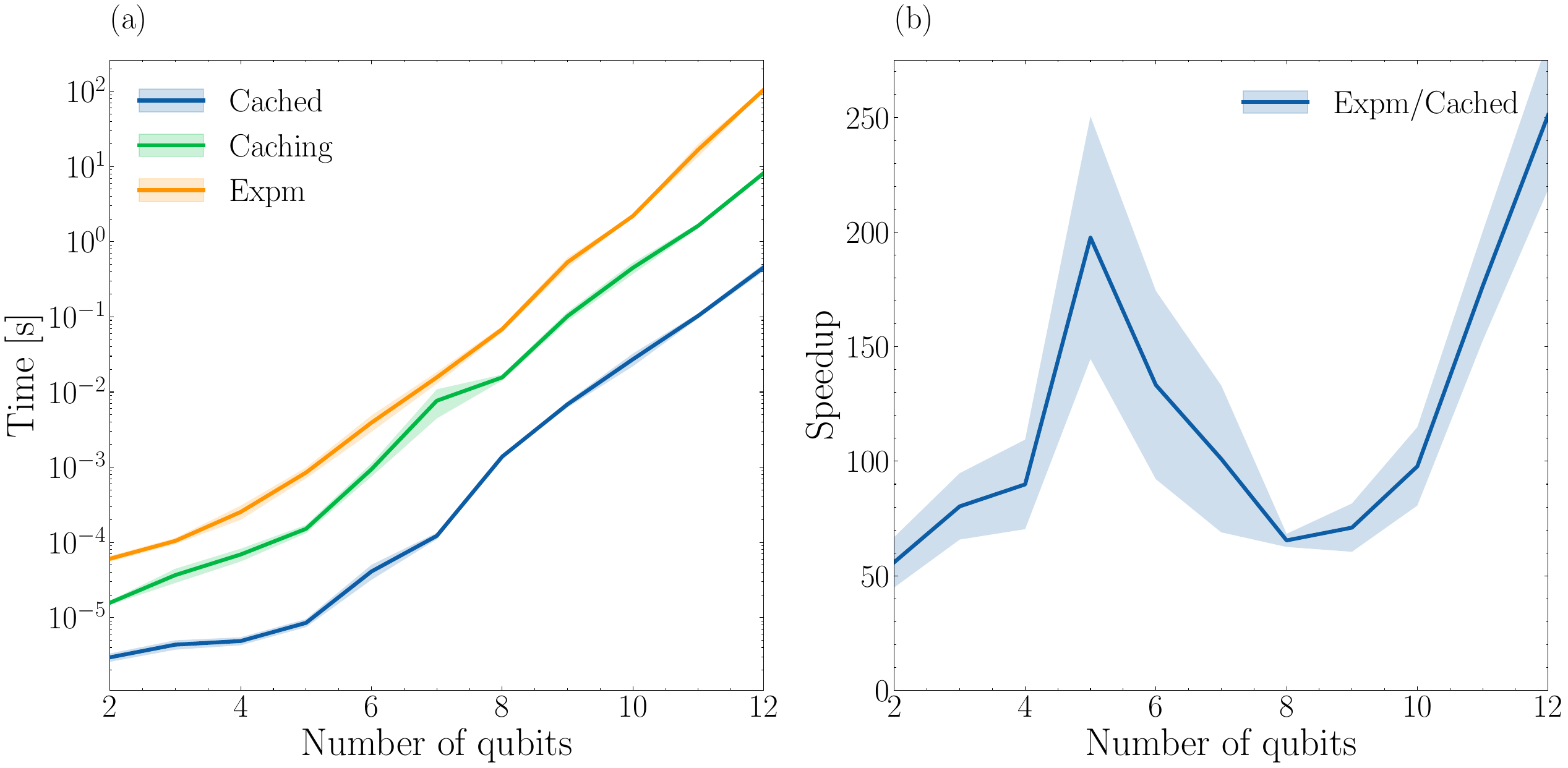}
\caption{(a) Walltime for MS gate construction with standard matrix exponentiation and with the analytical approach of this paper as a function of the number of qubits. The orange line corresponds to the function \textit{scipy.expm}, the blue and green lines represent the analytical approach during caching and after caching, respectively; (b) speedup of the new approach, once the parameter-indedependent matrices have been cached, as a function of the number of qubits. Here we see that the speedup is not constant as the number of qubits changes, but we have a maximum speedup around 5 qubits, which then decreases until we reach 8 qubits, and then increases again. This effect is most likely due to an underlying slowdown in the underlying fundamental operations, e.g., due to the processor cache being filled up quickly, so that access to the RAM becomes necessary. Moreover, our method is implemented only using NUMBA, i.e., compiled PYTHON code, whereas the exponential matrix function of SCIPY profits from underlying time-critical routines written in C, C++ and Fortran. It seems, however, that the speedup grows steadily for values larger than 8 qubits. \label{fig: times_and_speedups}}
\end{figure}

The approach used for $C_{xy}$ gate can be generalized to the MS gate. Due to its Hamiltonian being the square of the aforementioned Hamiltonian, the two gates share their eigenvectors, while the eigenvalues of the $C_{xy}$ gate are squared $\lambda_\text{XY,i}^2 = \lambda_\text{MS, i}$. This reduces the number of different eigenvalues from $n+1$ to $\lceil \frac{n}{2} \rceil +\mod(2+1, 2)$. The expansion in Eq.~\eqref{eq:xy_exp} is transformed into 
\begin{eqnarray}
\text{MS}(\phi, \theta) =& \frac{1}{2^n} v_\phi(\hat{C}) \odot \bigg(\sum_{i = 0}^{\lceil \frac{n}{2} \rceil} \exp(-i \lambda_\text{MS,i} \theta) \phantom{,} \overbrace{\left( \hat{D}_{\lambda_i} + \hat{D}_{\lambda_{n-i}} \right)}^{=\tilde{D}_i}\delta^{\text{floor}}_{i}\bigg).\label{eq:ms_exp}
\end{eqnarray}
where
\begin{align}
    \delta^{\text{floor}}_{i}  = \begin{cases}
     1 & \text{if $0 \leq i \leq \frac{n}{2}$}\\
     0 &  \text{otherwise}.
    \end{cases} 
\end{align}
As both expansions, Eq.~\eqref{eq:xy_exp} and  Eq.~\eqref{eq:ms_exp}, rely solely on operations local to the matrix elements, these operations are ideal for parallelisation.
In Fig.~\ref{fig: times_and_speedups} we show the improvement in terms of computation time (a) and speedup (b) of our method compared to standard matrix exponentiation as a function of the number of qubits : in (a) the orange line shows the computation time of the matrix exponential, whereas the green line shows our method implemented without caching the parameter-independent matrices, and the blue line our method again, but with cached matrices; (b) shows instead the ratio between the computation time of the cached method and standard matrix exponentiation. We see that we can reach a speedup between 50 and 250 for $n \leq 12$, which proves particularly useful in our quantum-circuit simulations: In fact, these require large numbers of cost function evaluations, due to the presence of both the reinforcement learning agent and the gradient-based optimization. 

\subsection{Gradients and Hessians}
The gradient of the $C_{xy}$ gate can then be computed analytically via 
\begin{eqnarray}
\pdv{\phi} C_{xy}(\phi, \theta) =& i \hat{C} \frac{1}{2^n} v_\phi(\hat{C}) \odot \sum_{i = 0}^n \exp(-i \lambda_\text{XY,i} \theta) \phantom{,} \hat{D}_{\lambda_i}, \label{eq:gradtheta_xy} \\
\pdv{\theta} C_{xy}(\phi, \theta) =& \frac{1}{2^n} v_\phi(\hat{C}) \odot \sum_{i = 0}^n -i \lambda_\text{XY,i} \exp(-i \lambda_\text{XY,i} \theta) \phantom{,} \hat{D}_{\lambda_i}. \label{eq:gradphi_xy}
\end{eqnarray}
For the derivative by $\phi$ the sum can be reused from the previous gate construction in Eq.~\eqref{eq:xy_exp} further reducing computational demands.
For the MS gate we find 
\begin{eqnarray}
\pdv{\phi} \text{MS}(\phi, \theta) =& i \hat{C}  \frac{1}{2^n} v_\phi(\hat{C}) \odot \bigg(\sum_{i = 0}^{\lceil \frac{n}{2} \rceil} \exp(-i \lambda_\text{MS,i} \theta) \phantom{,} \overbrace{\left( \hat{D}_{\lambda_i} + \hat{D}_{\lambda_{n-i}} \right)}^{=\tilde{D}_i}\delta^{\text{floor}}_{i}\bigg)\label{eq:gradtheta_ms}\\
\pdv{\theta} \text{MS}(\phi, \theta) =& \frac{1}{2^n} v_\phi(\hat{C}) \odot \bigg(\sum_{i = 0}^{\lceil \frac{n}{2} \rceil } -i \lambda_\text{MS,i} \exp(-i \lambda_\text{MS,i} \theta) \phantom{,} \overbrace{\left( \hat{D}_{\lambda_i} + \hat{D}_{\lambda_{n-i}} \right)}^{=\tilde{D}_i}\delta^{\text{floor}}_{i} \bigg).\label{eq:gradphi_ms}
\end{eqnarray}
Hessians can be derived direcly by applying the chain rule a second time in Eqs.~\eqref{eq:gradtheta_xy}-\eqref{eq:gradphi_ms}.

\begin{figure}[ht!]
\hspace{-0.7cm}
\centering
\includegraphics[width=\textwidth]{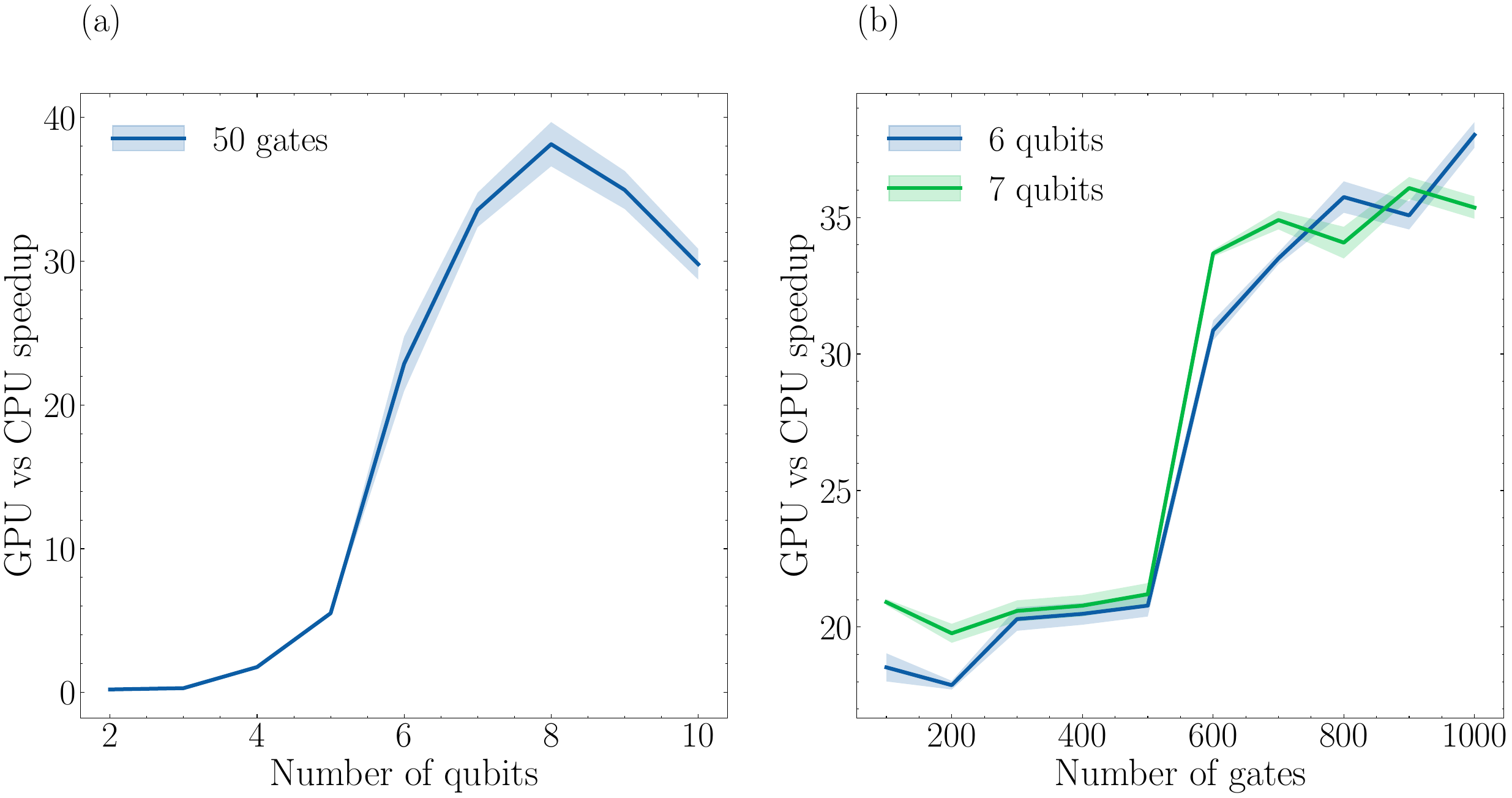}
\caption{Average speedup (30 shots) required to compute the gradient for: (a) the same circuit (50 gates) with increasing numbers of qubits. (b) a 6-qubit (blue) and 7-qubit (green) gate set for increasing number of random gates on the circuit. Shaded regions show the standard deviation for each point (vertical bars are connected together to form a poligon). The orange line represents the gradient compiled with NUMBA on CPU (Intel(R) Xeon(R) Gold 6240 CPU @ 2.60GHz), the second one the gradient compiled with JAX on GPU (NVIDIA Tesla V100S-PCIE-32GB). We see a slowdown in the speedup as we move past 8 qubits: this is probably due to computational overheads that do not profit from the GPU parallel calculations. In fact, we see that the slowdown is present only as we increase the Hilbert space dimension and not as we increase the number of gates.}
\label{fig:gradients}
\end{figure}

\section{Parameter-shift rules for ion gates}
The method for gate computation introduced above allows to express the derivative of a quantum cost function in terms of so-called parameter-shifts rules. These have been studied in the context of variational quantum circuit, quantum control and quantum machine learning. The gates contained in the gate set $Z_1(\theta), ..., Z_n(\theta)$, $C_{xy}(\theta, \phi)$, $\text{MS}(\theta, \phi)$.  
In general, expectation values with respect to parametrized quantum circuits will have the form
\label{appx:parametershift}
\begin{align}\label{eq:qcost}
   C(\boldsymbol{\alpha}) = \bra{\psi} V(\boldsymbol{\alpha}) \rho V^{\dagger}(\boldsymbol{\alpha}) \ket{\psi},
\end{align}
for a given quantum state $\ket{\psi}$, where $V(\boldsymbol{\alpha}) $ is a gate in the gateset defined in Section~\eqref{sec:gateset}. The cost function depends on a specific target operator $\rho$.

We first show that the cost function in Eq.~\eqref{eq:qcost} is equivalent to the cost function of Eq.~\eqref{eq:quantumcost_2}, as well as to the expectation value of the local and non-local Hilbert Schmidt circuit, which is given in Ref.~\cite{Khatri2019}:
\begin{align}
    C_{HST}(\boldsymbol{\alpha}) = 1 - \Tr{H (U \otimes V(\boldsymbol{\alpha})^*) \rho  (U^{\dagger} \otimes V(\boldsymbol{\alpha})^T)},
\end{align}
with $\rho = H = \ket{\phi_{+}}_{A, B} \bra{\phi_{+}}_{A,B}$ for the Hilbert Schmidt test, where $A$ and $B$ are the subystems for unitary $U$ and $V$ and
\begin{align}
    \ket{\phi_{+}}_{A, B} = \frac{1}{\sqrt{d}} \sum_{\boldsymbol{j}} \ket{\boldsymbol{j}_A} \otimes \ket{\boldsymbol{j}_B}.  
\end{align}
\begin{proof}
We show the equivalence for the Hilbert Schmidt test. The proof for the local Hilbert Schmidt test is analogous. \\
We see that
\begin{align*}
\Tr{H (UV^{\dagger} \otimes I) \rho (V \otimes U^{*})} =
\frac{1}{d} \sum_{\boldsymbol{j}} \Tr{\ket{\boldsymbol{j}_A} \bra{\boldsymbol{j}_A} U V^{\dagger} \ket{\boldsymbol{j}_A} \bra{\boldsymbol{j}_A}  VU^{\dagger}} \Tr{\ket{\boldsymbol{j}_B} \otimes \bra{\boldsymbol{j}_B}},
\end{align*}
which can be represented as a sum of squared amplitudes of with the same structure of \eqref{eq:qcost}.
\end{proof}
As a result, the cost functions considered here have to a common description, similar to Eq.~\eqref{eq:qcost}. We derive and comment the corresponding parameter-shift rule for each one of these gates. We consider here the shift of one single parameter at a time, i.e., a scalar rotation angle $\theta$. The gradient is constructed by shifting each parameter according to its own specific parameter-shift rule.
The derivatives of the function can be written as:
\begin{align}
    &\pdv{\theta}C(\theta) = \bra{\psi} \pdv{\theta} V(\theta) U V^{\dagger}(\theta) + V(\theta) U \pdv{\theta}  V^{\dagger}(\theta) \ket{\psi}.
\end{align}
Assuming the gate has a generator $V(\theta) = e^{i G \theta}$, where $G$ is a parameter-independent hermitian matrix, then we have:
\begin{align}
        &\pdv{\theta}C(\theta) = \bra{\psi} V(\theta) i[G,U] V^{\dagger}(\theta) \ket{\psi}.
\end{align} 
In the simplest case, we consider, e.g., $G = \sigma_z^{(i)}$ and obtain \cite{LiJun2017}:
\begin{align}\label{eq:pauli_shift}
    [\sigma_z^{(i)}, U] = e^{i \sigma_z^{(i)} \pi/2}Ue^{-i \sigma_z^{(i)} \pi/2} - e^{-i \sigma_z^{(i)} \pi/2}Ue^{i\sigma_z^{(i)} \pi/2},
\end{align}
which leads to
\begin{align}
    &\pdv{\theta}C(\theta) = C\left(\theta + \frac{\pi}{2}\right) - C\left(\theta - \frac{\pi}{2}\right),
\end{align}
a parameter-shift rule which is valid for the local $Z$-rotations. For the other two gates, we have to differentiate with respect to both the parameters $\theta$ and $\phi$. For $\theta$, applying Eq.~(15) in Ref.~\cite{Wierichs2022}, we have:
\begin{align}\label{eq:cxy_theta}
    \left. \pdv{\theta} C(\theta, \phi) \right|_{\theta=0} = \sum_{l=1}^{2n} \frac{(-1)^{l-1}}{2 \sin \left(\frac{2l-1}{2n}\pi \right) } C \left(\frac{2l-1}{2n}\pi, \phi \right).
\end{align}
By using the general decomposition of the $C_{xy}$ gates, parameter-shift rules can be obtained directly by studying how the eigenvalues and eigenvectors of the unitaries vary as a function of the continuous parameters. 
Since the $C_{xy}$ gate has $n + 1$ distinct eigenvalues we will need at most $2(n+1)$ samples to calculate the derivative -- see \cite{Wierichs2022}.
We observe that the matrices $\hat{V}_n(\phi)$ in Eq.~\eqref{eq:xy_decomposition_matrices} can be decomposed further in elementary gates, such as:
\begin{align}\label{eq:v_decomp}
    \hat{V}_n(\phi) = (\mathcal{P}(\phi)HX)^{\otimes n} = \mathcal{P}(\phi)^{\otimes n} H^{\otimes n} X^{\otimes n},
\end{align}
where
\begin{align}\label{eq:v_decomp_matrices}
    \mathcal{P}(\phi) = \begin{pmatrix} 1 & 0 \\ 0 & e^{i\phi} \end{pmatrix},  H=\frac{1}{\sqrt{2}}\begin{pmatrix} 1 & 1 \\ 1 & -1 \end{pmatrix}, X=\begin{pmatrix} 0 & 1 \\ 1& 0 \end{pmatrix}.
\end{align}

Let us first consider the $C_{xy}$ gate. The corresponding Hamiltonian has $n + 1$ different eigenvalues with energy $\lambda_i = 2i - n$ and $c_k(\phi) = v_k(\phi)^{\dagger} U v_k(\phi)$ are the overlaps between the target operator $U$ and the gate eigenvectors $v_k(\phi)$ of the gate. For the set of difference pairs we have $\lambda_i - \epsilon_j = 2(i-j) := 2l$. Then Eq.~\eqref{eq:cxy_theta} becomes

\begin{align}\label{eq:cxy_theta_simpl}
    \left. \pdv{\theta} C(\theta, \phi) \right|_{\theta=0} = \sum_{k=1}^n \sum_{l=1}^{2n} \frac{(-1)^{l-1}}{2 \sin \left(\frac{2l-1}{2n}\pi \right) } c_k(\phi) e^{2ik (\frac{2l-1}{2n})}.
\end{align}.

This expression can be simplified by evaluating the sum at $\theta=0$ with respect to the index $l$. 
\begin{align}\label{eq:cxy_theta_appx}
    \left. \pdv{\theta} C(\theta, \phi) \right|_{\theta=0} = \sum_{k=1}^{n} c_k(\phi) \frac{\csc \left(\frac{\pi }{2 n}\right) e^{\frac{\pi  i (k (4 n+2)-n)}{n}} \left(e^{\frac{2 \pi  i (n+1) (n-2 k)}{n}}-(-1)^{n+\frac{1}{n}}\right)}{2 n \left((-1)^n e^{4 \pi  i k}+e^{2 \pi  i n}\right)}.
\end{align}

For the MS gate we obtain instead the following expression:
\begin{align}
    \left. \pdv{\theta} C(\theta, \phi) \right|_{\theta=0} = \sum_{k=1}^{n} c_k(\phi) \left\{-\frac{\csc \left(\frac{\pi }{2 n}\right) e^{8 \pi  i k n-\frac{\pi  i (n-2 k)^2}{n}} \left((-1)^{n+\frac{1}{n}} e^{\frac{2 \pi  i (n+1) (n-2 k)^2}{n}}-1\right)}{2 n \left((-1)^n e^{2 \pi  i \left(4 k^2+n^2\right)}+e^{8 \pi  i k n}\right)}\right\}.
\end{align}
For the derivative of Eq.~\eqref{eq:qcost} with respect to $\phi$, we just have to consider the gate $\mathcal{P}(\phi)$ in Eq.~\eqref{eq:v_decomp}, whose single-qubit gate generator is given in Eq.~\eqref{eq:pgate}. We observe that due to the simple structure of the gate, the derivative of the $C_{x,y}$ gate with respect to $\phi$ is simply given by:
\begin{align}\label{eq:pgate}
    \mathcal{P}(\phi) =  \exp{\frac{i\phi}{2}(I_n - S_z)} = \exp{\frac{i\phi}{2}} \exp{-\frac{i\phi}{2} S_z},
\end{align}
where $S_z = \sum_{i=1}^n \sigma_z^{(i)}$ is the collective $Z$-operator acting on the entire qubit register. 
After inserting Eq.~\eqref{eq:pgate} into Eq.~\eqref{eq:qcost} for a $C_{xy}$ gate, and using the representation in Eq.~\eqref{eq:xy_diagonalisation}, we obtain
\begin{align}\label{eq:cost_phi}
    C(\theta, \phi) = \bra{\psi} \left( \exp{-\frac{i\phi}{2} S_z} B_{\text{XY}}(\theta)  \exp{\frac{i\phi}{2} S_z} U  \exp{-\frac{i\phi}{2} S_z} B_{\text{XY}}(\theta)^{\dagger}  \exp{\frac{i\phi}{2} S_z} \right) \ket{\psi},
\end{align}
where $B_{\text{XY}}(\theta) =  H^{\otimes n}X^{\otimes n} \left( \sum_{l=0}^{n} \exp{- i \lambda_l \theta} \ket{l} \bra{l} \right) X^{\otimes n} H^{\otimes n} $ -- the same equation holds for the $\text{MS}$ gate, one needs just to replace $\lambda_{\text{XY}}$ with the corresponding eigenvalues $\lambda_{MS}$.
Considering that $\mathcal{P}(\phi)$ has two eigenvalues, we can write
\begin{align}\label{eq:z_eigendecomp}
     \exp{-\frac{i\phi}{2} S_z} = e^{i\phi}\Pi^{(1)}_z + e^{-i\phi}\Pi^{(2)}_z, 
\end{align}
where $\Pi^{(1)}_z$ and $\Pi^{(2)}_z$ are the projectors on the respective degenerate eigenspaces. By using Eq.~\eqref{eq:z_eigendecomp} in Eq.~\eqref{eq:cost_phi}, we see that Eq.~\eqref{eq:cost_phi} has the following form
\begin{align}
    C(\theta, \phi) = b_0(\theta) + b_1(\theta) e^{-i \phi} + b_2(\theta) e^{i \phi} + b_3(\theta) e^{2 i \phi} + b_4(\theta) e^{- 2 i \phi},
\end{align}
where $b_i(\theta), i=0,1,2,3,4$ are $\phi$-independent coefficients. Thus, we can write the derivative of $C$ with respect to $\phi$ as
\begin{align}
    \pdv{\phi}C(\theta, \phi) =  - ib_1(\theta) e^{-i \phi} + ib_2(\theta) e^{i \phi} + b_3(\theta) 2i e^{2 i \phi} - 2i b_4(\theta) e^{- 2 i \phi},
\end{align}
which can be written as a linear combination of $C(\theta, \phi \pm \frac{\pi}{2})$ and $C(\theta, \phi \pm \frac{\pi}{4})$ cost functions:
\begin{align}
    \pdv{\phi}C(\theta, \phi) = \tilde{C}(\theta, \phi + \frac{\pi}{4}) - \tilde{C}(\theta, \phi - \frac{\pi}{4}) + (\frac{1}{\sqrt{2}} - \frac{1}{2})\tilde{C}(\theta, \phi - \frac{\pi}{2}) + (\frac{1}{\sqrt{2}} + \frac{1}{2})\tilde{C}(\theta, \phi + \frac{\pi}{2}).
\end{align}

\section{Algorithmic implementation details}
Our framework consists of two main parts: The agent (we consider a PS-LSTM agent with LSTM cells and a linear layer, but any RL agent is a viable option), which should learn to construct a proper representation of the quantum circuit and the optimizer, which has to be equipped with a proper gradient function. The gradient function is constructed according to the standard GRAPE procedure and using the gate representations for the $C_{xy}$ and the $\text{MS}$ gates given in Appendix \ref{sec:fast_gates}. We use and test two different versions of the gradient: The first one is compiled using NUMBA \cite{lam2015numba}, a library for fast python code, the second one employs JAX to allow execution on GPU. A comparison of the two gradient functions is given in Fig.~\ref{fig:gradients}. We observe that the GPU-based function allows for a certain speed-up, in particular as the number of gates on the circuit increases. The main advantage of NUMBA lies in a faster and more straightforward implementation of the parallel optimization runs with different seeds.

At time $t$ of the agent-environment interaction, the agent receives an input (percept) from the environment and outputs an action which is executed onto the environment.
The agent-environment interaction has the following structure. \\ \newline
\textbf{Action}: The action of the agent corresponds to placing one of the gates in the gate set (entangling or non-entangling) onto the quantum circuit.
The gate is represented by an integer $a \in {1, ...,  \vert A \vert}$. The array of all the integers chosen by the agent up to time $t$ forms the quantum circuit structure at time $t$. \\ \newline
\textbf{Percept}: As input to the RL agent, we generally use a one-hot encoding of the circuit. For a circuit of length $L$ and $\vert A \vert$ different gates, the percept $s_t \in \{0,1 \}^L \cross \{0,1 \}^{\vert A \vert}$ has entries equal to:
\begin{align}\label{eq:percept_standard}
    (s_t)_{i,j} = \begin{cases}
      1 & \text{if $i,j = a,t$}\\
      0 & \text{otherwise.}
    \end{cases}    
\end{align}
However, for the PS-LSTM agent (or other agents that are modified analogously), a more suitable input can also be used. Since the update of the internal state of the PS-LSTM agent follows the RL steps, the agent can just accept the input at time $t$ as a percept, since previous information about past agent-environment interactions is still processed by the the internal state of the LSTM network. Therefore, the percept becomes the one-hot encoded vector
\begin{align}\label{eq:percept_lstm}
    (s_t)_{j} = \begin{cases}
      1 & \text{if $j = a$}\\
      0 & \text{otherwise.}
\end{cases}    
\end{align}
This percept only gives the agent partial information about the internal state of the environment, which turns the problem from an MDP (Markov Decision Process) into a so-called POMDP problem (Partially Observable Markov Decision Process) \cite{Preti2020}.
Other agents than those derived by PS may use different inputs, based on their specific convergence properties in dealing with different environments. In general, if the agent can accept a recurrent or autoregressive network as a policy, the percept given in Eq.~\eqref{eq:percept_lstm} may be used. It could be, however, that the algorithm needs to be modified appropriately to function with this type of percept. The first type of input has been used for the simulation of UCC operators, whereas the second one has been implemented in all the other simulations. In general, both percepts lead to similar results, but the second one should be preferred for the PS-LSTM architecture and most importantly does not scale with the size of the circuit, leading therefore to faster forward passages in the policy network. \\ \newline
\textbf{Reward and curriculum}:
Throughout the development of this manuscript, several reward systems were tested. 
\begin{figure*}
\includegraphics[width=\textwidth]{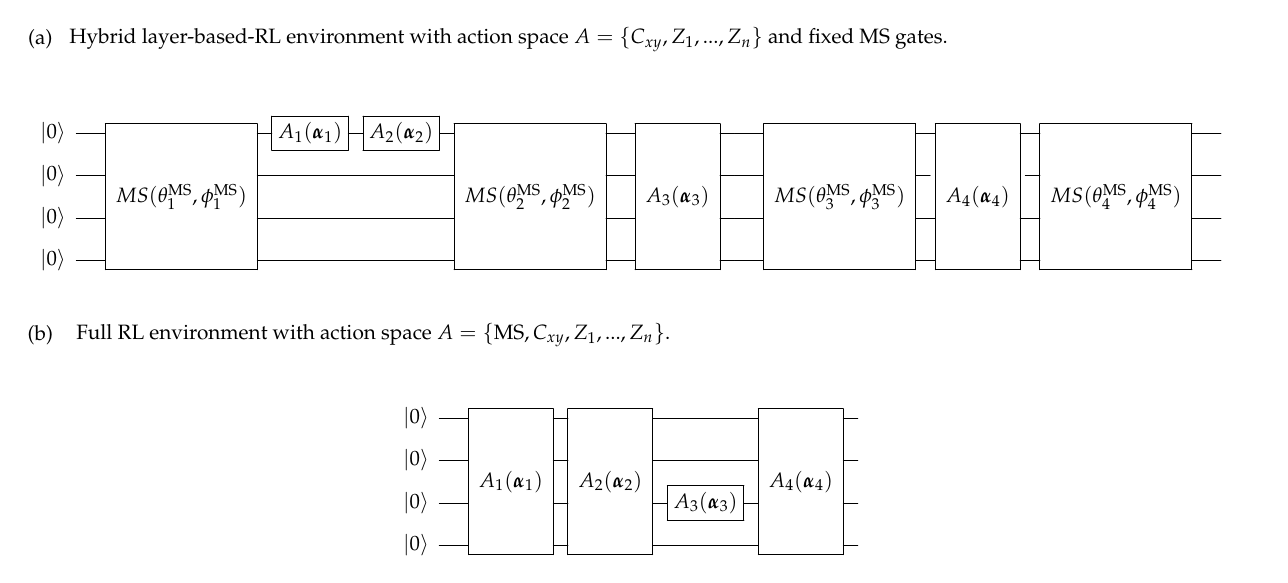}
\caption{Quantum circuit representation of two possible RL environments for quantum circuit optimization. The first environment (a) resembles the structure of the layer-based compilation, that is, the entangling MS gates are fixed and the agent can place rotations on the circuit between the entangling layers. The second one (b) has no pre-defined circuit structure but rather leaves the agent completely free to place any gate withing the gate set on the quantum circuit, with only one additional simplification: two gates of the same type placed immediately one after the other are automatically merged to form one single gate. This is done to prevent the agent from getting stuck in loops, i.e., local minima, where it keeps choosing the same gate over and over again. In general, the first circuit reduces the size of the action space and therefore the possible shapes of the corresponding cost function, hence reducing exploration in favour of a more standardized search.}
\label{fig:layer-non-layer}
\end{figure*}
In general, we used a two-step reward that assigns a smaller reward value when the cost function minimum falls below the actual curriculum threshold $\epsilon_t$ and a larger reward value when the cost function minimum falls below the global target threshold $\epsilon_{\text{min}}$. Both curriculum thresholds can be adjusted depending on the gate synthesis problem to be tackled.
The episode terminates when the cost function minimum in a given time step falls below the threshold $\epsilon_t$ or when the maximal length of the circuit per episode, $L_{\text{max}}$, is reached. The RL training terminates upon reaching the maximum number of episodes $E_{\text{max}}$. The reward scheme helps to progressively increase the fidelity throughout training without allowing for too long circuits.
The threshold is then lowered as episodes progress based on previous rewards obtained by the agent. In our implementation, we lower the threshold when it has been surpassed by the agent at least 500 times using the following scheme:
\begin{align}\label{eq:threshold_update_appendix}
    \epsilon_{t+1} = \begin{cases}
      \epsilon_{\text{min}} + \frac{1}{2}(\epsilon_{\text{min}} - \epsilon_t) & \text{if $\epsilon_{\text{min}} \leq \epsilon_t \leq 1$}\\
      \epsilon_{\text{min}} & \text{if $\epsilon_t \leq \epsilon_{\text{min}}$}\\
      1 & \text{otherwise}
    \end{cases}    
\end{align}
\newline
\textbf{Environments}: We propose two different types of agent-circuit interaction. In the first type -- see Fig.~\ref{fig:layer-non-layer} (a) --, which is more similar to layer-based compilation, the agent only places rotations on the circuit, while the entangling gates are fixed in position. While the episode progresses, more entangling gates are added to the environment, until a maximum $L_{\text{max}}$ of gates on the circuit is reached. The circuit is also simplified automatically by the environment, in such a way that if the agent places the same two gates on the circuit one after the other, they are reduced to one single gate. This is implemented in order to prevent the agent from getting trapped in local minima where it places the same gate over and over again on the circuit, therefore reducing the necessary training time. In the second type -- see Fig.~\ref{fig:layer-non-layer} (b) --, entangling gates are also given as a possible action on the environment. This second type of environment leaves more room for exploration, but it is also more challenging for the agent. The simulations presented in this work were realized by using only the environment that allows for completely free gate placement (so both the rotation gates as well as the entangling gates). \\ \newline
\textbf{Agents}: In order to test the effectiveness of our algorithm, we employ different agents for testing. We consider the standard state-of-the-art PPO algorithm \cite{pytorch_minimal_ppo} and the two versions of PS with deep energy-based neural networks PS-DEBN and PS-LSTM. We observe that PPO performs slightly worse than PS-LSTM, but better than PS-FNN, but is probably due to the non-recurrent version of the PPO algorithm implemented. Due to the large action and percept space, we did not include standard PS in the comparisons, since this would require to store large $h$-matrices for each percept-action transition, leading to slow computation and eventually memory overflow. This algorithm can however be used in circuit synthesis problems with a small number of gates and qubits. \\ \newline
\textbf{Optimizers}:
As an unconstrained optimizer, in this work we only consider the L-BFGS-B algorithm, as it is implemented in SCIPY \cite{2020SciPy-NMeth}. The number of iterations of the optimizers for each RL interaction is set to 100. Both optimizers can run a given number of optimization attempts in parallel with different seeds (a technique usually referred to as random restart), which should prevent the optimizer from getting trapped in local minima. \\
\newline
\textbf{Simulations}:
In this work we consider three different groups of simulations: the synthesis of a $3$-qubit Toffoli gate on a $3$-qubit and $4$-qubit circuit, the synthesis of UCC (Unitary Coupled Cluster) operators and the synthesis of the unitary of the XXZ Hamiltonian with varying Hamiltonian parameters. In the case of the XXZ Hamiltonian and the Toffoli gate, the simulation is realized on CPUs (72 Intel(R) Xeon(R) Gold 6240 CPU @ 2.60GHz), whereas the simulation of the UCC operators is performed on GPUs (4 NVIDIA A100 Tensor Core GPU with 40 GB). This means, the library used to compute cost functions and gradients in order to optimize them is NUMBA for the CPU-based simulations and JAX for GPU-based ones. The reason is that NUMBA allows for faster parallelization with the JOBLIB library that proves difficult to achieve with JAX, thereby enabling us a better exploration of the cost function landscape for small numbers of qubits, whereas JAX is significantly faster for larger numbers of qubits. In particular, we set the number of optimization runs per RL iteration to 10 when we employ the NUMBA version, whereas we use only 1 when we employ the code using JAX. The curriculum threshold was kept to $\epsilon_{\text{min}} = 10^{-2}$. We used two-layered LSTM networks implementing the policy given in Eq.~\eqref{eq:lstm_policy} and 128 neurons, a training batch size of 64, a learning rate of 0.01, a curriculum update window of 500. The agent is trained with a replay-memory upon finding a gate sequence with infidelity lower than the curriculum threshold, and also every 100 agent-environment interactions if the replay memory is large enough. The target network is updated every 50 agent-environment interactions. The number of iterations of the optimizer in the hybrid RL-continuous simulations is fixed to 100. The temperature parameter $\beta$ of the softmax distibution that parametrized the PS-LSTM policy -- see Eq.~\eqref{eq:psh_policy} -- is annealed from $\beta=10^{-3}$ to  $\beta=1$ according to a linear schedule based on the number of episodes, in order to force the agent, towards the end of the training, to consider shorter and shorter circuits, thereby reducing the exploration. Other parameters vary based on the simulation considered. The data and hyperparameters can be found in \href{https://github.com/franz3105/RL_Ion_gates}{\url{https://github.com/franz3105/RL_Ion_gates}}. 
\\
\newline
\textbf{Comparison with BQSKit}:
BQSKit is a quantum circuit compilation library developed by the Berkeley National Laboratory \cite{BQSKit}. It supports compilation of both discrete and variational circuits with different algorithms, such as tree-search with BFS, qsearch, etc. and allows for the implementation of custom gate sets. For comparison, we implement the trapped-ion gate set used in this work -- see also Eqs.~\eqref{eq: ms gate set}-\eqref{eq: ms gate set3} -- and run the compilation using the QSearch algorithm implemented in the aforementioned library, which is based on multiple tree traversal search algorithm such as the $A^*$ algorithm \cite{Hart1968} or the Dijkstra algorithm \cite{Hodler2019}. The results of these runs for the Toffoli gate are given in Table~\ref{tab:bqskit}: we see that the $A^*$ and Dijkstra routines give similar solutions, whereas the greedy heuristic proves significantly worse overall. The compilation routine offered by BQSKit, while significantly faster than the RL method, is unable to converge to the same compact solution discovered by the PS-LSTM algorithm. 

\begin{table}[]
	\centering
	\begin{tabular}{|c|c|c|c|c|}
		\hline
		Gate type & $A^*$ search & Dijkstra & Greedy & our method \\
		\hline
		MS gate & 5 & 3 & 4 & 3 \\
		$C_{x,y}$ gate & 3 & 4 & 94 & 4 \\
		$Z$ gate & 27 & 24 & 297 & 3 \\
		\hline
		\hline
		Depth & 17 & 15 & 113 & 10 \\
		Runtime & 17 s  & 63 s & 161 s & 1.6 h  \\
		\hline 
	\end{tabular}
	\caption{Result of the compilation of the 3-qubit Toffoli gate with BQSKit \cite{BQSKit} using the gate set given by Eqs.~\eqref{eq: ms gate set}-\eqref{eq: ms gate set3} and three of the given search heuristics: $A^*$, Dijkstra, and greedy. The threshold is set to $\epsilon = 10^{-2}$. We see that the compilation uses a considerable amount of $Z$ gates and a larger amount of collective gates than the RL algorithm. We see, however, that a standard compiler can be significantly faster than a RL-based search with 10000 episodes and it is therefore possible to apply further pruning methods and iterative optimization loops on top of the main compilation algorithm.}
	\label{tab:bqskit}
\end{table}

\end{document}